\documentclass[twoside]{article}
\usepackage{geometry}
\usepackage{Shorthands}
\usepackage[round]{natbib}

\title{Performance-Robustness Tradeoffs in Adversarially Robust Linear-Quadratic Control}

\author{Bruce D.\ Lee\thanks{Equal contribution} , Thomas T.C.K.\ Zhang\footnotemark[1] , Hamed Hassani, and Nikolai Matni}

\date{Department of Electrical and Systems Engineering, University of Pennsylvania}

\newcounter{cdc}
\begin{document}

\maketitle
 \begin{abstract}
    While $\calH_\infty$ methods can introduce robustness against worst-case perturbations, their nominal performance under conventional stochastic disturbances is often drastically reduced. Though this fundamental tradeoff between nominal performance and robustness is known to exist, it is not well-characterized in quantitative terms. Toward addressing this issue, we borrow from the increasingly ubiquitous notion of adversarial training from machine learning to construct a class of controllers which are optimized for disturbances consisting of mixed stochastic and worst-case components. We find that this problem admits a stationary optimal controller that has a simple analytic form closely related to suboptimal $\calH_\infty$ solutions.
    We then provide a quantitative performance-robustness tradeoff analysis, in which system-theoretic properties such as controllability and stability explicitly manifest in an interpretable manner. This provides practitioners with general guidance for determining how much robustness to incorporate based on a priori system knowledge. We empirically validate our results by comparing the performance of our controller against standard baselines, and plotting tradeoff curves.
\end{abstract}
\section{Introduction}

Modern dynamical systems, from mobile robotics to power plants, require controllers that are simultaneously fast, efficient, and robust. Many control schemes attempt to simultaneously achieve these desiderata by combining them into a single objective function and optimizing it. By combining these distinct goals into a single objective, a natural tradeoff arises. A controller optimized for speed and efficiency may perform poorly in the face of unmodeled phenomena. 
For instance, Linear-Quadratic Gaussian (LQG) controllers (a special case of $\calH_2$ controllers) explicitly prioritize nominal performance by penalizing the expectation of a quadratic function of the state and input. It has been shown, however, that such controllers can be extraordinarily fragile to small perturbations of the dynamics \citep{doyle1978guaranteed}. Replacing the LQG objective with one that considers the response of the system to worst-case perturbations of the dynamics and external disturbances results in robust control methods, such as $\calH_\infty$ methods. Such controllers are provably robust, however they tend to be overly conservative. 

Toward achieving a balance between the performance of nominal and robust controllers, various approaches have been introduced, most notably mixed $\calH_2/\calH_\infty$ methods. However, the resulting controllers are often complicated to express and compute \citep{carsten1995mixed}, and lack \textit{a priori} quantitative guarantees on how much nominal performance must be given up so as to attain a desired robustness level. Toward addressing these issues, we take inspiration from the notion of adversarial robustness from machine learning \citep{madry2017towards, carlini2017towards} in order to formulate a controller synthesis that balances performance and robustness. The goal of adversarial robustness is to minimize the expected error under the presence of worst-case norm-bounded perturbations to the data, where the perturbations can depend on the underlying stochasticity of the problem, such as the data distribution and additive noise. We consider an analogous adversarially robust state feedback control problem where we aim to minimize the expected LQ cost where the process noise is composed of two components: a zero-mean stochastic noise component and a norm-bounded adversarial perturbation. We show that the solution to this problem admits a closed-form expression in terms of discrete algebraic Riccati equation (DARE) solutions, which in turn allows for novel quantitative performance-robustness tradeoff bounds in which system parameters manifest in a  natural and interpretable way. 

\subsection{Contributions}


\ifnum\value{cdc}>0{
Toward analyzing the adversarially robust control problem we propose, we first show that when viewed through the lens of dynamic games \citep{basar1991hinf}, adversarially robust LQ control relates to a control problem introduced in \cite{doyle1989optimal}. We show that the optimal solution to the state feedback version of this problem is given by a central static suboptimal $\calH_\infty$ controller, with suboptimality level $\gamma$ depending on both the stochastic noise statistics and the budget given to the adversary. Furthermore, both the worst-case adversary and the corresponding optimal controller can be computed from the solution of a DARE.   

We quantify the performance-robustness tradeoffs of adversarially robust LQ control, both analytically and empirically, and show a clear and interpretable dependence on underlying system theoretic properties such as controllability and stability. In particular, we show that the cost gap incurred by the adversarially robust controller in the nominal setting over the nominal controller is upper bounded by $O(\sigma_w^2 \gamma^{-4} \nu^{-1})$, where $\sigma_w^2$ is the covariance of the additive noise distribution, $\gamma$ is the suboptimality level of the suboptimal $\calH_\infty$ controller, and $\nu$ is the smallest singular value of the controllability gramian. On the other hand, the cost gap is lower bounded by $\Omega\paren{\sigma_w^2 \gamma^{-4} \eta^2 }$, where $\eta = \norm{W_\infty(A+BK_\star, I)}$ is the largest singular value of the controllability gramian for closed loop system under the nominal LQ controller with disturbances as the input. These results show that systems with uniformly good controllability have small performance-robustness tradeoffs, while those that have a highly controllable mode in the nominal closed-loop system (when viewing disturbances as inputs) lead to large performance-robustness tradeoffs.
} \else {
\begin{itemize}
\item Toward analyzing the adversarially robust control problem we propose, we first show that when viewed through the lens of dynamic games \citep{basar1991hinf}, adversarially robust LQ control relates to a control problem introduced in \cite{doyle1989optimal}. We show that the optimal solution to the state feedback version of this problem is given by a central static suboptimal $\calH_\infty$ controller, with suboptimality level $\gamma$ depending on both the stochastic noise statistics and the budget given to the adversary. Furthermore, both the worst-case adversary and the corresponding optimal controller can be computed from the solution of a DARE.   
\item We quantify the performance-robustness tradeoffs of adversarially robust LQ control, both analytically and empirically, and show a clear and interpretable dependence on underlying system theoretic properties such as controllability and stability. In particular, we show that the cost gap incurred by the adversarially robust controller in the nominal setting over the nominal controller is upper bounded by $O(\sigma_w^2 \gamma^{-4} \nu^{-1})$, where $\sigma_w^2$ is the covariance of the additive noise distribution, $\gamma$ is the suboptimality level of the suboptimal $\calH_\infty$ controller, and $\nu$ is the smallest singular value of the controllability gramian. On the other hand, the cost gap is lower bounded by $\Omega\paren{\sigma_w^2 \gamma^{-4} \eta^2 }$, where $\eta = \norm{W_\infty(A+BK_\star, I)}$ is the largest singular value of the controllability gramian for closed loop system under the nominal LQ controller with disturbances as the input. These results show that systems with uniformly good controllability have small performance-robustness tradeoffs, while those that have a highly controllable mode in the nominal closed-loop system (when viewing disturbances as inputs) lead to large performance-robustness tradeoffs.
\end{itemize}
} \fi

\subsection{Related Work}

The mixed stochastic/worst-case problem that we consider is not the only way to strike a balance between the performance of stochastic and robust control methods. Most closely related is \cite{doyle1989optimal}, which considers disturbances composed of both a bounded power component, and a bounded power spectrum component, and approaches the problem through a deterministic viewpoint. 
\cite{paganini1993set} considers a set description of disturbances that also interpolates between $\calH_2$ and $\calH_\infty$ approaches.
\cite{glover1988state} provide the class of all stabilizing controllers subject to a $\calH_\infty$ norm constraint, while \cite{bernstein1988LQG} and \cite{rotea1991h2} consider minimizing $\calH_2$ objective subject to a $\calH_\infty$ constraint. These methods often lack a simple closed-form stationary solution. Other recent work attempts to reduce the conservatism of robust control by taking a regret-minimization approach \citep{goel2020regret, goel2021competitive, hazan2020nonstochastic}.  None of the aforementioned methods provide a characterization of the cost incurred by enforcing robustness.



Analogous recent work in the machine learning community has analyzed performance-robustness tradeoffs in adversarially robust learning, including precise characterizations of the generalization errors of standard versus adversarially trained models under the random features model \cite{javanmard2020precise, hassani2022curse}, and ``no free lunch'' theorems for obtaining adversarially robust models \cite{tsipras2018robustness, dohmatob2019generalized, yin2019rademacher}.  The successful characterization of such performance-robustness tradeoffs in machine learning motivates the control objective we consider. However, the theoretical results from this area are largely intended for the supervised learning setting and do not immediately apply to our setting. \cite{al2020accuracy} proves the existence of performance-robustness tradeoffs in control but does not characterize them quantitatively. The extension of adversarial robustness results in machine learning to various control problems is an active area of research \citep{lee2021adversarial, zhang2021adversarially, havens2022revisiting}. 

\textbf{Notation:} The Euclidean norm of a vector $x \in \R^n$ is denoted by $\norm{x}$. The spectral norm of a matrix $A \in \R^{m\times n}$ is denoted $\norm{A}$, and the Frobenius norm of $A \in \R^{m \times n}$ is denoted $\norm{A}_F$. The spectral radius of a square matrix $A \in \R^{n \times n}$ is denoted $\rho(A)$. A symmetric, positive semidefinite (psd) matrix $A = A^\top$ is denoted $A \succeq 0$. Similarly $A \succeq B$ denotes that $A-B$ is positive semidefinite. A symmetric positive definite (pd) matrix $A=A^\top$ is denoted $A \succ 0$.
A sequence of vectors $x_t \in \R^n$ defined for $t \geq 0$ will be denoted by $\mathbf{x} = \curly{x_t}_{t \geq 0}$. The $\ell^2$ signal-norm of a sequence is denoted by $\norm{\mathbf{x}}_{\ell^2}$.
The solutions to a discrete Lyapunov equation and discrete algebraic Riccati equation are denoted $\dlyap(A, Q)$ and $\dare(A, B, Q, R)$, respectively.

\ifnum\value{cdc}>0{
\Bruce{Include reference to arxiv version, say that all proofs are available there. Remove all references to the appendix.}
}\else{}\fi
\section{Adversarially Robust Linear-Quadratic Control} \label{sec: advrobust lq control}

Consider a fully observed discrete-time linear-time-invariant (LTI) system with state  disturbances composed of both stochastic and adversarial components: let $x_t \in \R^n$ be the system state, $u_t \in \R^m$ the input, $w_t \in \R^n$ and $\delta_t \in \R^n$ the stochastic and adversarial components of the process disturbance, respectively. The initial condition $x_0$ and stochastic component of the process noise $w_t$ are assumed to be i.i.d.\ zero-mean with covariance matrices $\Sigma_0,\;\Sigma_w$, respectively, and $\Ex\brac{x_0 w_t^\top}= 0$ for all $t \geq 0$. The LTI system is then defined by the following equation:
\begin{equation}\label{eq:ss}
    \begin{aligned}
     x_{t+1} &= Ax_t + Bu_t + w_t + \delta_t.
\end{aligned}
\end{equation}
We assume that the adversarial perturbation sequence $\boldsymbol{\delta}$ is causal, i.e., it can depend only on the states, inputs, and stochastic noise up to the current timestep; in particular, it must be a measureable function of the randomness $x_0, w_{0:t}$.
We let $Q \succeq 0$ be a weight matrix for the state cost, and $R \succ 0$ be a weight matrix for the input cost. We consider the infinite horizon objective 
\[
    \limsup_{T\to\infty} \frac{1}{T} \Ex_{\mathbf{w}, x_0} \brac{x_T^\top Q x_T + \sum_{t=0}^{T-1} x_t^\top Q x_t + u_t^\top R u_t}
\]
subject to the dynamics \eqref{eq:ss}. 
If the adversarial perturbation is set to zero, $\boldsymbol{\delta} = 0$, then the above objective is the nominal LQR problem. If the stochasticity is set to zero, $\boldsymbol{w}, x_0 =0$, and $\boldsymbol{\delta}$ are worst-case perturbations with average power bounded by $\varepsilon$, the above objective is the $\calH_\infty$ problem. When both stochastic noise and worst-case perturbations are present, we define the resulting control task as the adversarially robust LQR problem.
We denote the three corresponding objectives by $\NC$, $\RC$, and $\AC$ respectively:
\ifnum\value{cdc}>0{
\begin{align}
    &V(T,K,Q,R,\mathbf{x}):=x_T^\top Q x_T + \sum_{t=0}^{T-1} x_t \paren{Q + K^\top R K}x_t \nonumber \\
    \begin{split}
    &\NC(K) := \limsup_{T \to \infty} 
    \frac{1}{T} \Ex_{\mathbf{w}, x_0}  \brac{V(T,K,Q,R,\mathbf{x})}, \\
    &\quad\quad\quad\quad\quad \textnormal{s.t. } x_{t+1} = (A+BK)x_t + w_t \label{eq:nominal_cost}
    \end{split}\\
    \begin{split}
    &\RC(K) := \limsup_{T \to \infty} \frac{1}{T} \max_{\substack{\boldsymbol{\delta} \mbox{ causal} \\ \norm{\boldsymbol{\delta}}_{\ell_2}^2 \leq T \varepsilon}} V(T,K,Q,R,\mathbf{x}), \label{eq:robustcost} \\ 
    &\quad\quad\quad\quad\quad \textnormal{s.t. } x_{t+1} = (A+BK)x_t +  \delta_t,\; x_0 = 0
    \end{split}\\
    &\AC(K)  := \limsup_{T \to \infty} \frac{1}{T} \Ex_{\mathbf{w}, x_0}  \brac{\max_{\substack{\boldsymbol{\delta} \mbox{ causal} \\ \norm{\boldsymbol{\delta}}_{\ell_2}^2 \leq T \varepsilon}}  V(T,K,Q,R,\mathbf{x})}, \label{eq:hardadversarialLQRobjective} \nonumber \\
    &\quad\quad\quad\quad\quad \textnormal{s.t. }x_{t+1} = (A+BK)x_t + w_t + \delta_t. 
\end{align}
}\else{
\begin{align}
     \label{eq:nominal_cost}
    \NC(K) &= \limsup_{T \to \infty} 
    \frac{1}{T} \Ex_{\mathbf{w}, x_0}  \brac{ x_T^\top Q x_T + \sum_{t=0}^{T-1} x_t \paren{Q + K^\top R K}x_t }, \;\; x_{t+1} = (A+BK)x_t + w_t \\
    \label{eq:robustcost}
    \RC(K) &= \limsup_{T \to \infty} \frac{1}{T} \max_{\substack{\boldsymbol{\delta} \mbox{ causal} \\ \norm{\boldsymbol{\delta}}_{\ell_2}^2 \leq T \varepsilon}} x_T^\top Q x_T + \sum_{t=0}^{T-1} x_t^\top (Q+K^\top R K) x_t, \;\; x_{t+1} = (A+BK)x_t +  \delta_t,\; x_0 = 0 \\
    \label{eq:hardadversarialLQRobjective} 
    \AC(K)  &= \limsup_{T \to \infty} \frac{1}{T} \Ex_{\mathbf{w}, x_0}  \brac{\max_{\substack{\boldsymbol{\delta} \mbox{ causal} \\ \norm{\boldsymbol{\delta}}_{\ell_2}^2 \leq T \varepsilon}} x_T^\top Q x_T + \sum_{t=0}^{T-1} x_t^\top (Q+K^\top R K) x_t }, \;\; x_{t+1} = (A+BK)x_t + w_t + \delta_t 
\end{align}
}\fi
The adversarial budget $\norm{\boldsymbol{\delta}}_{\ell^2}^2 \leq T\varepsilon$ is chosen such that the instance-wise adversarial budget satisfies $\norm{\delta_t}^2 \leq \varepsilon$ on average.
We note that our attention is restricted to static state feedback control policies $u_t = Kx_t$, as they are known to be optimal for $\NC$ and $\RC$ \citep{zhou1996robust}. We will show, as a consequence of Lemma~\ref{lem:adversarial LQR sol} and Theorem~\ref{thm:adversarial controller solution}, that they are in fact also optimal for $\AC$.
In order to ensure that there exists a stabilizing controller, and that minimizing either $\NC$ or $\RC$ provides a stabilizing controller, we make the standard assumption that $(A,B,Q^{1/2})$ is stabilizable and detectable \citep{zhou1996robust}.
Under this assumption, there exists a stabilizing state feedback control law $u_t = K x_t$ minimizing $\NC$, where
\begin{align}
    \label{eq:lqr}
    K = -(R + B^\top P B)^{-1} B^\top P A 
\end{align}
and $P$ is the solution to the Discrete Algebraic Riccati Equation (DARE) 
\begin{align}
    \label{eq:nominal DARE}
    P = Q + A^\top P A - A^\top P B(R + B^\top P B)^{-1} B^\top P A.
\end{align}
A solution minimizing $\RC$ may be found using Theorem 13.3.3 of \cite{hassibi1999}.

The remainder of this section is devoted to finding a controller minimizing $\AC$.  
Inspired by minimax dynamic games \citep{basar1991hinf}, we first find a controller which minimizes a soft-constrained version of the adversarial cost \eqref{eq:hardadversarialLQRobjective}:
\ifnum\value{cdc}>0{
\begin{equation}
\begin{aligned}
    \label{eq:adversarialLQRobjective}
     \limsup_{T \to \infty} 
    \frac{1}{T} \Ex \bigg[\max_{\boldsymbol{\delta} \mbox{ causal}} x_T^\top Q x_T &+ \sum_{t=0}^{T-1} x_t Q x_t \\
    &+ u_t^\top R u_t  - \gamma^2 \delta_t^\top \delta_t\bigg].
\end{aligned}
\end{equation}
} \else {
\begin{align}
    \label{eq:adversarialLQRobjective}
     \limsup_{T \to \infty} 
    \frac{1}{T} \Ex \brac{\max_{\boldsymbol{\delta} \mbox{ causal}} x_T^\top Q x_T + \sum_{t=0}^{T-1} x_t Q x_t + u_t^\top R u_t - \gamma^2 \delta_t^\top \delta_t}.
\end{align}
} \fi
The following lemma provides necessary and sufficient conditions for the existence of a stabilizing controller which minimizes the above objective. Similar statements in continuous time may be extracted from a more general result found in \cite{doyle1989optimal}. 

\begin{lemma}
    \label{lem:adversarial LQR sol} A controller attaining a finite value for objective \eqref{eq:adversarialLQRobjective} exists if and only if there exists a solution to the following discrete algebraic Riccati equation   
    \ifnum\value{cdc}>0{
    \begin{align}
         \label{eq:modified DARE}
        P = \dare\paren{A, \bmat{B & I}, Q, \bmat{R & 0\\ 0 & -\gamma^2 I}}
    \end{align}
    }\else{
         \begin{align}
         \label{eq:modified DARE}
        P = Q + A^\top P A - A^\top P \bmat{B & I} \paren{\bmat{B & I}^\top P \bmat{B & I} +\bmat{R & 0\\ 0 & -\gamma^2 I}}^{-1} \bmat{B & I}^\top P A
        \end{align}
    }\fi
    satisfying $0 \preceq P \prec\gamma^2 I$.
    As long as the above condition is true, 
    \begin{enumerate}
        \item The controller $u_t = K x_t$ with $K$ given by
        \begin{equation}
        \label{eq:suboptimal adversarial controller}
        \begin{aligned}
            K &= -(R+B^\top M B)^{-1} B^\top M A, \\
            M &= P + P(\gamma^2 I - P)^{-1} P.
        \end{aligned}
        \end{equation}
        satisfies $\rho(A+BK) < 1$, and minimizes the objective  \eqref{eq:adversarialLQRobjective}. \label{num:1}
        \item \label{num:2} The optimal adversarial perturbation under the controller $u_t=Kx_t$ is given by 
        \begin{equation}
        \label{eq:adversarial perturbation}
        \begin{aligned}
            \Delta &= (\gamma^2 I - P)^{-1} P \\
            \delta_t &=  \Delta((A+BK) x_t + w_t).
        \end{aligned} 
        \end{equation}
        \item  The objective value \eqref{eq:adversarialLQRobjective} achieved under controller \eqref{eq:suboptimal adversarial controller} and adversary \eqref{eq:adversarial perturbation} is  $\trace(M \Sigma_w)$.
    \end{enumerate}
\end{lemma}


The solution approach for the above problem follows that in \cite{basar1991hinf} for minimax games. 
The finite horizon version of the problem is solved by defining a saddle point cost-to-go, then recursing backwards in time, solving both an unconstrained concave maximization problem and convex minimization problem at each time step to determine the optimal adversarial perturbation and control input. The causality of $\boldsymbol \delta$ is necessary in the recursion
to pull $\delta_t$ out of an expectation over future noise terms. Taking the limit as the horizon tends to infinity provides the steady state controller and adversary in the above lemma statement.  It should be noted that in contrast to most adversarially robust machine learning problems, adversarially robust LQR provides a closed form expression for the adversarial perturbation. 

We now return our attention to the objective \eqref{eq:hardadversarialLQRobjective}.  We show via strong duality that the hard-constrained problem may be solved by sequentially solving the soft-constrained problem using Lemma~\ref{lem:adversarial LQR sol}. Note that in contrast to the solution approach to minimize $\RC$, dualizing the constraint in $\AC$ results in an optimal dual variable $\gamma(\varepsilon)$ that is a random variable. Therefore, it is nontrivial to exchange the order of the minimization over the dual variable with the expectation (see Appendix~\ref{appendix: advlqr thm proof} for details).  We propose Algorithm~\ref{alg: AdvLQR} to minimize the adversarial cost AC \eqref{eq:hardadversarialLQRobjective}.  





\begin{algorithm}
\label{alg:advLQR}
\caption{Computing Adversarially Robust Controller: \textbf{\texttt{AdvLQR}}$(A,B,Q,R,\varepsilon, \gamma_{LB}, \gamma_{UB}, \texttt{tol})$} 
\label{alg: AdvLQR}

\begin{algorithmic}[1]
\State \textbf{Input:} State matrices $A,B$, cost matrices $Q,R$, adversary budget $\varepsilon > 0$, binary search bounds $\gamma_{LB}<\gamma_{UB}$, tolerance $\texttt{tol}$.\;
\State // \emph{Do binary search on $\gamma \in [\gamma_{LB}, \gamma_{UB}]$ to find optimal adversary with average power $\varepsilon > 0$}
\While{$\gamma_{UB}-\gamma_{LB}\geq \texttt{tol}$}
\State $\gamma = (\gamma_{UB} +\gamma_{LB})/2$
\State Compute $P, M, K, \Delta$ at level $\gamma$ via \eqref{eq:modified DARE}-\eqref{eq:adversarial perturbation}
\State $G = \dlyap((A+BK)^\top \Delta, \Delta \Sigma_w \Delta)$.
\If{$\trace(G (A+BK)^\top (I+\Delta)^2 (A+BK) + \Delta \Sigma_w \Delta) < \varepsilon$ \eqref{eq: expected adversary power}}
\State $\gamma_{UB} = \gamma$
\Else 
\State$\gamma_{LB} = \gamma$
\EndIf
\EndWhile
\State \textbf{Output:} Adversarially robust LQR controller $K$, adversary gain $\Delta$, optimal value of \eqref{eq:hardadversarialLQRobjective} $\trace(M \Sigma_w)+\gamma^2 \varepsilon$.
\end{algorithmic}

\end{algorithm}




\begin{theorem}
    \label{thm:adversarial controller solution}
    Suppose $(A,B, Q^{1/2})$ is stabilizable and detectable. Let $\gamma_\infty$ be the optimal closed-loop $\calH_\infty$ gain of system \eqref{eq:ss}. For any $\varepsilon > 0$, let $\gamma_{UB}>\gamma_\infty$ satisfy the following condition: if $P, M, K, \Delta$ satisfy equations \eqref{eq:modified DARE}-\eqref{eq:adversarial perturbation} at level $\gamma_{UB}$ and $G := \dlyap((A+BK)^\top \Delta, \Delta\Sigma_w\Delta)$, then for the disturbance $\delta_t$ in equation \eqref{eq:adversarial perturbation},
    \ifnum\value{cdc}>0{
    \begin{equation}
    \begin{aligned}\label{eq: expected adversary power}
        \lim_{t \to \infty} \Ex\brac{\delta_t^\top \delta_t} = \trace(&G (A+BK)^\top (I+\Delta)^2 \\
        & \cdot(A+BK) + \Delta \Sigma_w \Delta) < \varepsilon.
    \end{aligned}
    \end{equation}
    }\else{
    \begin{align}\label{eq: expected adversary power}
        \lim_{t \to \infty} \Ex\brac{\delta_t^\top \delta_t} = \trace(G (A+BK)^\top (I+\Delta)^2 (A+BK) + \Delta \Sigma_w \Delta) < \varepsilon.
    \end{align}
    }\fi
    Under these conditions, the output of Algorithm~\ref{alg: AdvLQR} \textnormal{\texttt{AdvLQR}}$(A,B,Q,R,\varepsilon, \gamma_\infty, \gamma_{UB}, \textnormal{\texttt{tol}})$ satisfies
    \begin{enumerate}
        \item The control policy $u_t = K x_t$ minimizes $\AC(K)$. 
        \item The optimal adversarial perturbation in equation \eqref{eq:hardadversarialLQRobjective} under the optimal policy is $\delta_t = \Delta \paren{(A+BK)x_t + w_t}$ and satisfies $\lim_{t \to \infty} \Ex\brac{\delta_t^\top \delta_t} \leq \varepsilon$.
        \item The minimum value for the adversarial cost \eqref{eq:hardadversarialLQRobjective} is given by $\trace(M\Sigma_w)+\gamma^2 \varepsilon$. 
    \end{enumerate}


    %
\end{theorem}

We make the interesting observation from Theorem~\ref{thm:adversarial controller solution} that while the nominal LQR controller is independent of the noise statistics, the adversarially robust controller output by Algorithm~\ref{alg: AdvLQR} \textit{is} dependent on the noise statistics through the optimal choice of $\gamma$.


\begin{remark}
We note that the same results used to solve for the optimal controller minimizing $\AC(K)$ may be adapted to evaluate $\AC(K)$ under an arbitrary stabilizing controller $K$. In particular, observe that $(A+BK, 0)$ is stabilizable, and $((Q+K^\top R K)^{1/2}, A+BK)$ is detectable. 
Then considering the closed loop system and cost matrices $(A', B', Q', R'):= (A+BK, 0, Q+K^\top RK, 0)$, we see that 
Lemma~\ref{lem:adversarial LQR sol} implies that under controller $u_t = Kx_t$, equation \eqref{eq:adversarialLQRobjective} evaluates to $\trace(M\Sigma_w)$, where $M = P + P(\gamma^2 I - P)^{-1}P$, and $P=\dare(A+BK, I, Q+K^\top RK, -\gamma^2 I)$. Similarly,  Theorem~\ref{thm:adversarial controller solution} tells us that for any $\varepsilon > 0$, with properly selected $\gamma_{UB}, \gamma_{LB}$ and $\textnormal{\texttt{tol}}$, $\AC(K)$ may be determined via Algorithm~\ref{alg: AdvLQR} as \textnormal{\texttt{AdvLQR$(A+BK,0,Q+K^\top R K,0,\varepsilon, \gamma_{LB}, \gamma_{UB}, \texttt{tol})$}}. 
\end{remark}


\section{Performance-Robustness Tradeoff Bounds}
\label{s: performance robustness tradeoffs}



This section presents the main results of the paper: we study the tradeoffs that arise in adversarial control by investigating the interplay between the two objectives \eqref{eq:nominal_cost} and \eqref{eq:hardadversarialLQRobjective}. This section will focus on bounding the tradeoffs analytically. The next section will validate these analytical bounds by empirically examining tradeoff curves for particular systems. 

We begin by considering the gap between the nominal controller and the $\gamma$-adversarially robust controller when  applied in the nominal setting, i.e., we seek to bound the gap $\NC(K_\gamma) - \NC(K_\star)$, where $K_\gamma$ is the $\gamma$-adversarially robust controller given by Lemma~\ref{lem:adversarial LQR sol}, and $K_\star$ is the nominal LQR controller given by equation \eqref{eq:lqr}. 
Let $P_\star$ be the nominal DARE solution given by the solution to equation \eqref{eq:nominal DARE}, and let $P_\gamma$ be the solution to equation \eqref{eq:modified DARE}. Also define 
\begin{align}
    \label{eq:Mgamma}
    M_\gamma := P_\gamma + P_\gamma(\gamma^2I - P_\gamma)^{-1} P_\gamma.
\end{align}

Given an arbitrary stabilizing linear feedback controller $K$, 
results from \cite{mania2019certainty} and \cite{fazel2018global} allow us to characterize the gap in the cost between $K$ and the optimal LQR controller $K_\star$ as
\ifnum\value{cdc}>0{
$\NC(K) - \NC(K_\star) 
    =\; \trace(\Sigma(K) (K-K_\star)^\top (R + B^\top P_\star B)(K- K_\star))
$
}
\else{
\begin{align}
    \label{eq: fazel result}
    \NC(K) - \NC(K_\star) = \trace(\Sigma(K) (K-K_\star)^\top (R + B^\top P_\star B)(K- K_\star))
\end{align}
}\fi
where $\Sigma(K) = \dlyap(A+BK, \Sigma_w)$ is the steady state covariance of the closed loop system under controller $K$. The following bounds on the gap $\NC(K_\gamma) - \NC(K_\star)$ then follow immediately:
\ifnum\value{cdc}>0{
\begin{align}
    \label{eq:cost gap lower/upper bound}
    \begin{split}
    &\sigma_{\min}(\Sigma_w) \sigma_{\min}(R + B^\top P_\star B) \norm{K_\gamma - K_\star}_F^2 \\
    \leq\; &\NC(K_\gamma) - \NC(K_\star) \\
    \leq\; &\norm{\Sigma(K_\gamma)} \norm{R + B^\top P_\star B}  \norm{K_\gamma - K_\star}_F^2.
    \end{split}
\end{align}
}
\else{
\begin{align}
    \label{eq:cost gap lower bound}
    \NC(K_\gamma) - \NC(K_\star) &\geq \sigma_{\min}(\Sigma_w) \sigma_{\min}(R + B^\top P_\star B) \norm{K_\gamma - K_\star}_F^2 \\
    \label{eq:cost gap upper bound}
    \NC(K_\gamma) - \NC(K_\star) &\leq \norm{\Sigma(K_\gamma)} \norm{R + B^\top P_\star B}  \norm{K_\gamma - K_\star}_F^2.
\end{align}
}\fi
We have therefore reduced the task of upper and lower bounding the cost gap between the $\gamma$-adversarially robust controller and the nominal LQR controller in the nominal setting to directly bounding the gap between the two controllers. Recalling that $\norm{K_\gamma - K_\star}^2_F \leq \min\curly{m,n} \norm{K_\gamma - K_\star}^2$, we use  
the following lemma to bound the difference $\norm{K_\gamma - K_\star}$ in terms of the difference between the solutions to the  corresponding adversarial and nominal DAREs. 

\begin{lemma}(Adapted from Lemma 2 of \cite{mania2019certainty})
    \label{lem: controller gap to DARE gap}
    Suppose that $f_1(u; x) = \frac{1}{2} u^\top R u + \frac{1}{2}(Ax + Bu)^\top M (Ax+Bu)$ and $f_2(u; x) = \frac{1}{2} u^\top R u + \frac{1}{2}(Ax + Bu)^\top P (Ax+Bu)$ with $M \succeq P$. Furthermore, for any $x$, let $u_i = K_i x = \argmin_u f_i(u,x)$. Then we have that 
    \ifnum\value{cdc}>0{
    \begin{align*}
        \frac{\norm{B^\top (M - P) (A+BK_2)}}{\norm{R + B^\top M B)}} &\leq \norm{K_1 - K_2} \\ 
        &\leq \frac{\norm{B^\top (M - P) (A+BK_2)}}{\sigma_{\min}(R + B^\top P B)}.
    \end{align*}
    }\else{
    \[
        \frac{\norm{B^\top (M - P) (A+BK_2)}}{\norm{R + B^\top M B)}} \leq \norm{K_1 - K_2} \leq \frac{\norm{B^\top (M - P) (A+BK_2)}}{\sigma_{\min}(R + B^\top P B)}.
    \]
    }\fi
\end{lemma}


For the rest of this paper, we assume for simplicity that $\Sigma_w = \sigma_w^2 I$. We also define $\gamma_\infty$ as the minimum $\calH_\infty$ norm for the closed loop system, i.e., the smallest value of $\gamma$ for which the conditions of Lemma~\ref{lem:adversarial LQR sol} hold. Similarly, we define $\tilde \gamma_\infty$ as the $\calH_\infty$ norm of the closed loop system under the nominal LQR controller. Additionally, we define the $\ell$-step controllability gramian as $W_{\ell}(A,B): = \sum_{t=0}^\ell A^t BB^\top (A^t)^\top$. If $\rho(A) < 1$ we define the controllability gramian as $W_\infty(A,B):=\lim_{\ell \to \infty} W_\ell(A,B)$.



\subsection{Upper Bound}\label{sec: upper bound}



Applying Lemma~\ref{lem: controller gap to DARE gap} with $P =P_\star$ and $M = M_\gamma$  reduces our goal to bounding the spectrum of $M_\gamma - P_\star$. From the definition of $M_\gamma$ \eqref{eq:Mgamma}, we can write $
    \norm{P_\star - M_\gamma} \leq \norm{P_\star - P_\gamma} + \frac{\norm{P_\gamma}^2}{\gamma^2 - \norm{P_\gamma}}.$
For $\gamma > \gamma_\infty$ we have $P_\gamma \prec P_{\gamma_\infty} \prec \gamma_\infty^2 I$ (Lemma~\ref{lem: gamma ordering}), and thus $
    \norm{P_\star - M_\gamma} \leq \norm{P_\star - P_\gamma} + \frac{\gamma_\infty^4}{\gamma^2 - \gamma_\infty^2},$
reducing our task to bounding $\norm{P_\star-P_\gamma}$, the gap between solutions to the $\gamma$-adversarial and nominal DAREs. 
Toward bounding the norm difference of solutions to DAREs, we show that the closed loop dynamics under the adversary $\delta_t$ can be expressed as perturbations of the nominal system matrices. In particular, recall that for the a noiseless adversarial LQR instance at level $\gamma > 0$, the adversary can be represented as $
\delta_t =  \paren{\gamma^2 I - P_\gamma }^{-1} P_\gamma \paren{Ax_t + Bu_t},$
such that we may write $x_{t+1} = Ax_t + Bu_t + \delta_t = \tilde{A}x_t + \tilde{B}u_t$, for $\tilde A := \paren{I + (\gamma^2 I -P_\gamma)^{-1}P_\gamma}A$ and $\tilde B :=\paren{I + (\gamma^2 I -P_\gamma)^{-1}P_\gamma}B $. This allows us to bound the gaps $||\tilde{A} - A||,\; ||\tilde{B}-B||$ in terms of $\gamma$. This is formalized in Lemma~\ref{lem: eps to gamma conversion} in Appendix~\ref{appendix: upper bound}.




By bounding the gap between system matrices in the adversarial setting and in the nominal setting, we derive perturbation bounds on the gap %
$\norm{P_\gamma-P_\star}$ between the adversarial and nominal DARE solutions, which ultimately leads to the following upper bound on $\NC(K_\gamma) - \NC(K_\star)$.

\begin{theorem}
\label{thm: nominal cost gap upper bound}
Suppose $(A,B, Q^{1/2})$ is controllable and detectable. Define the condition number $\kappa(Q, R) := \frac{\max\curly{\sigma_{\max}(Q), \sigma_{\max}(R)}}{\min\curly{\sigma_{\min}(Q), \sigma_{\min}(R)}}$, 
$\tau(A, \rho) := \sup\curly{\norm{A^k}\rho^{-k}: k \geq 0}$,
and $\beta := \max\curly{1, \frac{\gamma_{\infty}^2}{\gamma^2 - \gamma_\infty^2} \tau\paren{A, \rho} + \rho}$, where $\rho > \rho(A)$. Furthermore, 
let $\ell$ be any natural number $1 \leq \ell \leq n$. For $\gamma > 0$ satisfying
\ifnum\value{cdc}>0{
\begin{align*}
    \gamma^2 \geq \gamma_\infty^2 + &\frac{3}{2} \ell^{3/2} \beta^{\ell - 1} \sigma_{\min}(W_\ell(A,B))^{-1/2} \tau(A, \rho)^2 \\
    &\cdot \paren{\norm{B} + 1} \max\curly{\norm{A}, \norm{B}}  \gamma_\infty^2,
\end{align*}
}
\else{
\begin{align*}
    \gamma^2 \geq \gamma_\infty^2 + \frac{3}{2} \ell^{3/2} \beta^{\ell - 1} \sigma_{\min}(W_\ell(A,B))^{-1/2} \tau(A, \rho)^2 \paren{\norm{B} + 1} \max\curly{\norm{A}, \norm{B}}  \gamma_\infty^2,
\end{align*}
}\fi
the following upper bound holds:
\ifnum\value{cdc}>0{
\begin{align*}
    &\NC(K_\gamma) - \NC(K) \\
    &\leq O(1) \sigma_w^2 \Big(\frac{\gamma_\infty^4}{\gamma^2 - \gamma_\infty^2}\Big)^2 m \ell^{5} \beta^{4(\ell - 1)} \norm{A + BK_\star}^2 \tau(A, \rho)^6 \\
    &\quad \cdot \paren{1 + \sigma_{\min}(W_\ell(A,B))^{-1/2}}^2 \norm{W_\infty (A + BK_\gamma, I)}  \\
    &\quad \cdot \frac{\norm{R + B^\top P_\star B}}{\sigma_{\min}\paren{R + B^\top P_\star B}^2} \kappa(Q,R)^2 \norm{B}^2\paren{\norm{B} + 1}^4  \norm{P_\star}^2.
\end{align*}
}
\else{
\begin{align*}
    \NC(K_\gamma) - \NC(K) &\leq O(1) \sigma_w^2 \Big(\frac{\gamma_\infty^4}{\gamma^2 - \gamma_\infty^2}\Big)^2 m \ell^{5} \beta^{4(\ell - 1)} \paren{1 + \sigma_{\min}(W_\ell(A,B))^{-1/2}}^2  \\
    &\quad \cdot \norm{A + BK_\star}^2 \norm{W_\infty (A + BK_\gamma, I)} \tau(A, \rho)^6 \\
    &\quad \cdot \frac{\norm{R + B^\top P_\star B}}{\sigma_{\min}\paren{R + B^\top P_\star B}^2} \kappa(Q,R)^2 \norm{B}^2\paren{\norm{B} + 1}^4  \norm{P_\star}^2.
\end{align*}
}\fi

\end{theorem}
As $\gamma \to \infty$, our upper bound decays to $0$ as expected, since the adversarial controller converges to the nominal controller in the limit. However, the steepness of this cost gap is affected by system properties such as the minimum singular value of the $\ell$-step controllability gramian, where poor controllability causes the upper bound to increase. We note that in contrast to the perturbation gap requirements in \cite{mania2019certainty}, our condition on the perturbation gap via lower bounds on $\gamma$ are much less stringent. We only require a lower bound on $\gamma$ to guarantee that the controllability of the adversarially perturbed system $(\tilde{A}, \tilde{B})$ is on the same order of magnitude as that of the nominal system $(A,B)$. 

\subsection{Lower Bound}\label{s: p-r tradeoffs lower bounds}
Applying Lemma \ref{lem: controller gap to DARE gap} with $M = M_\gamma$ and $P = P_\star$, we conclude that
\ifnum\value{cdc}>0{
\begin{align}
    \label{eq: general controller lb}
    \norm{K_\gamma - K_\star} \geq \frac{\norm{B^\top (M_\gamma-P_\star) } \sigma_{\min}(A+BK_\star) }{\norm{R + B M_\gamma B}}.
\end{align}
}\else{
\begin{align}
    \label{eq: general controller lb}
    \norm{K_\gamma - K_\star} &\geq \frac{\norm{B^\top (M_\gamma-P_\star) (A+BK_\star) }}{\norm{R + B M_\gamma B}} \geq \frac{\norm{B^\top (M_\gamma-P_\star) } \sigma_{\min}(A+BK_\star) }{\norm{R + B M_\gamma B}}.
\end{align}}\fi
Next, we add and subtract a particular DARE solution to the $M_\gamma-P_\star$ term in the above bound. Specifically, for $\gamma \geq \tilde \gamma_\infty$, we let $\tilde P_\gamma = \dare(A+BK_\star, I, Q, -\gamma^2 I)$, and note that $x^\top \tilde P_\gamma x$ represents the cost of applying controller $K_\star$ in the adversarial setting at level $\gamma$ starting from state $x$ with $w_t = 0$ for all $t \geq 0$. Adding and subtracting $\tilde P_\gamma$ in the lower bound in \eqref{eq: general controller lb} yields
\ifnum\value{cdc}>0{
\begin{align}
    \nonumber
    &\norm{K_\gamma - K_\star}\\
    &\geq \frac{\norm{B^\top (M_\gamma-\tilde P_\gamma + \tilde P_\gamma - P_\star) }\sigma_{\min}(A+BK_\star) }{\norm{R + B M_\gamma B}} \\
    &\geq  \frac{\sigma_{\min}(A+BK_\star)}{\norm{R + B M_\gamma B}} \cdot\paren{\norm{B^\top (P_\gamma\Delta_\gamma + \tilde P_\gamma - P_\star)  }-\norm{B^\top (\tilde P_\gamma - P_\gamma)}}.
    \label{eq:lb on controller gap}
\end{align}
}
\else{
\begin{align}
    \nonumber
    \norm{K_\gamma - K_\star}
    &\geq \frac{\norm{B^\top (M_\gamma-\tilde P_\gamma + \tilde P_\gamma - P_\star) }\sigma_{\min}(A+BK_\star) }{\norm{R + B M_\gamma B}} \\
    &\geq  \frac{\sigma_{\min}(A+BK_\star)}{\norm{R + B M_\gamma B}} \paren{\norm{B^\top (P_\gamma(\gamma^2 I - P_\gamma)^{-1} P_\gamma + \tilde P_\gamma - P_\star)  }-\norm{B^\top (\tilde P_\gamma - P_\gamma)}}.
    \label{eq:lb on controller gap}
\end{align}
}\fi
To obtain a lower bound on the above expression, we can upper and lower bound the spectra of $\tilde P_\gamma - P_\gamma$ and $\tilde P_\gamma - P_\star$, respectively. To upper bound the spectrum of $\tilde P_\gamma - P_\gamma$, we apply a result similar to equation \eqref{eq: fazel result} for the noiseless setting. We may lower bound the spectrum of $\tilde P_\gamma - P_\star$ by writing each as the solution to a Lyapunov equation and observing that their difference may also be written as the solution to a Lyapunov equation. This is stated formally and proven in Lemma~\ref{lem: spectral UB+LB}. Combining the above leads to the following theorem.


\begin{theorem}
\label{thm:lower bound}
\ifnum\value{cdc}>0{Define $\tilde{\Delta}_\gamma := I+(\gamma^2 I - \tilde P_\gamma)^{-1} \tilde P_\gamma$. }\fi Suppose $(A,B,Q^{1/2})$ is stabilizable and detectable, $\gamma \geq \tilde{\gamma}_\infty$, and
\ifnum\value{cdc}>0{
\begin{align*}
    \gamma^2 &\geq \sigma_{\min}(P_\star) + \frac{1}{2}\sigma_{\min}(P_\star)^2\frac{\norm{B^\top W_\infty(A+BK_\star, I) }}{\norm{R + B^\top M_\gamma B}} \\
    &\cdot\norm{B^\top  W_\infty\paren{\tilde{\Delta}_\gamma (A +B K_\star), I}} \sigma_{\min}(A+BK_\star)^2.
\end{align*}
}
\else{
\begin{align*}
    \gamma^2 &\geq \sigma_{\min}(P_\star) \\
    &+ \frac{1}{2}\sigma_{\min}(P_\star)^2\frac{\norm{B^\top W_\infty(A+BK_\star, I) }}{\norm{R + B^\top M_\gamma B}} \norm{B^\top  W_\infty\paren{ \paren{I+(\gamma^2 I - \tilde P_\gamma)^{-1} \tilde P_\gamma} (A +B K_\star), I}} \sigma_{\min}(A+BK_\star)^2.
\end{align*}
}\fi
Then the following lower bound holds:
\ifnum\value{cdc}>0{
\begin{align*}
    &\NC(K_\gamma) - \NC(K_\star) \\
    &\geq \frac{\sigma_w^2}{2} \bigg( \frac{\sigma_{\min}(P_\star)^2}{\gamma^2 - \sigma_{\min}(P_\star)} \bigg)^2 \frac{\sigma_{\min}(R + B^\top P_\star B)}{\norm{R + B^\top M_\gamma B}^2} \\
    &\quad\cdot\norm{B^\top W_\infty(A+BK_\star, I) }^2 \sigma_{\min}(A+BK_\star)^2.
\end{align*}
}
\else{
\begin{align*}
    \NC(K_\gamma) - \NC(K_\star) &\geq \frac{\sigma_w^2}{2} \bigg( \frac{\sigma_{\min}(P_\star)^2}{\gamma^2 - \sigma_{\min}(P_\star)} \bigg)^2 \frac{\sigma_{\min}(R + B^\top P_\star B)}{\norm{R + B^\top M_\gamma B}^2} \norm{B^\top W_\infty(A+BK_\star, I) }^2 \sigma_{\min}(A+BK_\star)^2.
\end{align*}
}\fi

\end{theorem}
Keeping the nominal system fixed, both the upper and lower bounds decay at a rate $\gamma^{-4}$. We note that instead of the $\ell$-step controllability gramian that manifests in the upper bound, we have instead the system parameter $W_{\infty}(A+BK_\star, I)$, which can be seen as a ``disturbability gramian'', which is the controllability gramian of the closed loop system under controller $K_\star$ with disturbances as inputs. That is, a large $\norm{B^\top W_\infty(A+BK_\star, I) }$ implies that the nominal closed-loop system is quantifiably more controllable by the distrubance input, hence more susceptible to adversarial disturbances of fixed energy.

To apply the upper and lower bounds to the adversarial setting with fixed adversarial budget $\varepsilon$, one may find the optimal corresponding $\gamma$ via Algorithm~\ref{alg: AdvLQR}.

\section{Numerical Experiments} \label{sec: numerical experiments}

We now empirically study the trends suggested by our adversarially robust controller synthesis and the subsequent performance-robustness tradeoff bounds.

\ifnum\value{cdc}>0{
\begin{figure}
    \label{fig:tradeoff curves}
    \centering
    \subfloat[Tradeoff curves]{\label{fig:tradeoff curves}\includegraphics[width=0.4\textwidth]{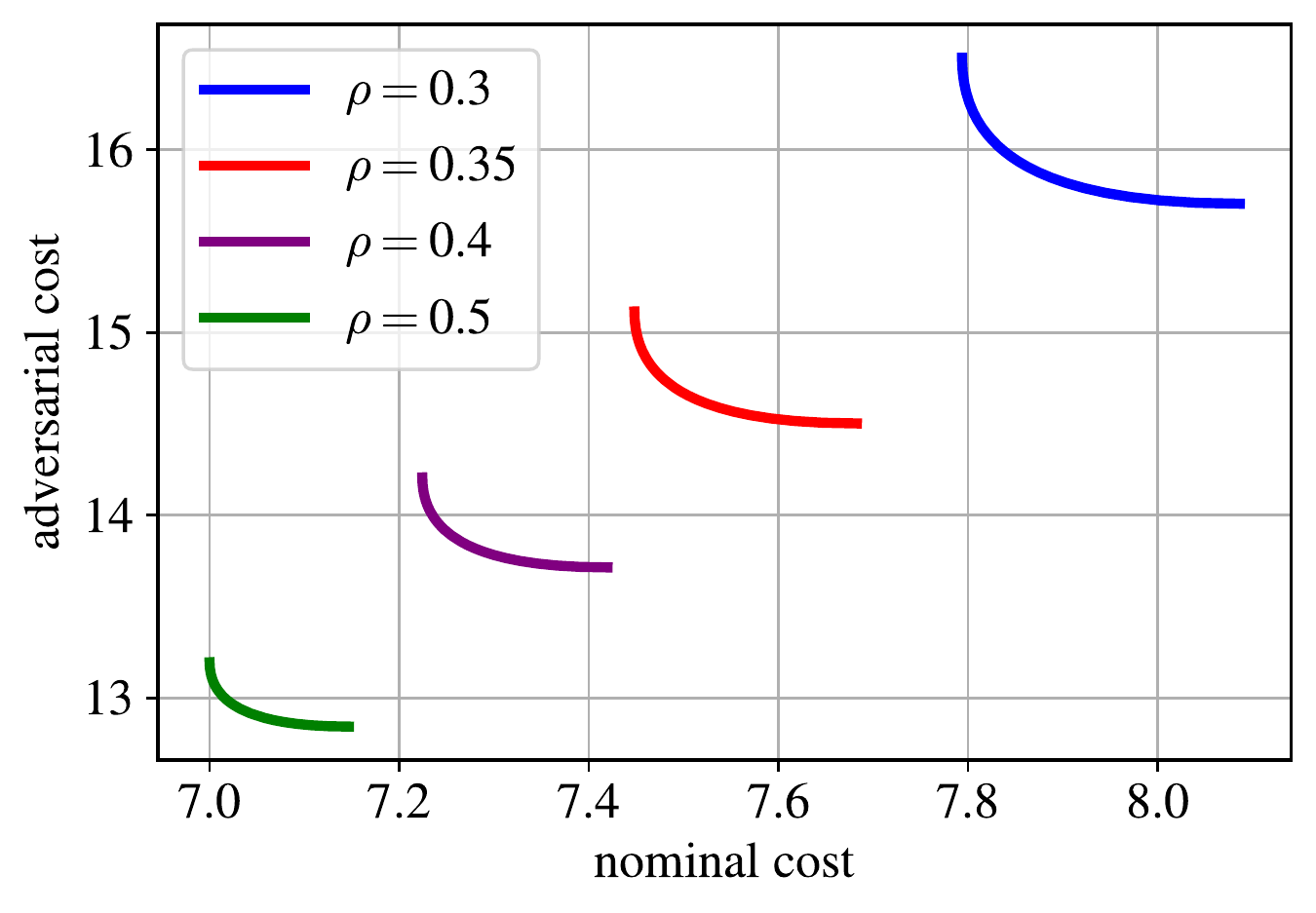}}  \\
    \subfloat[Tradeoff envelope]{\label{fig:tradeoff envelope}\includegraphics[width=0.4\textwidth]{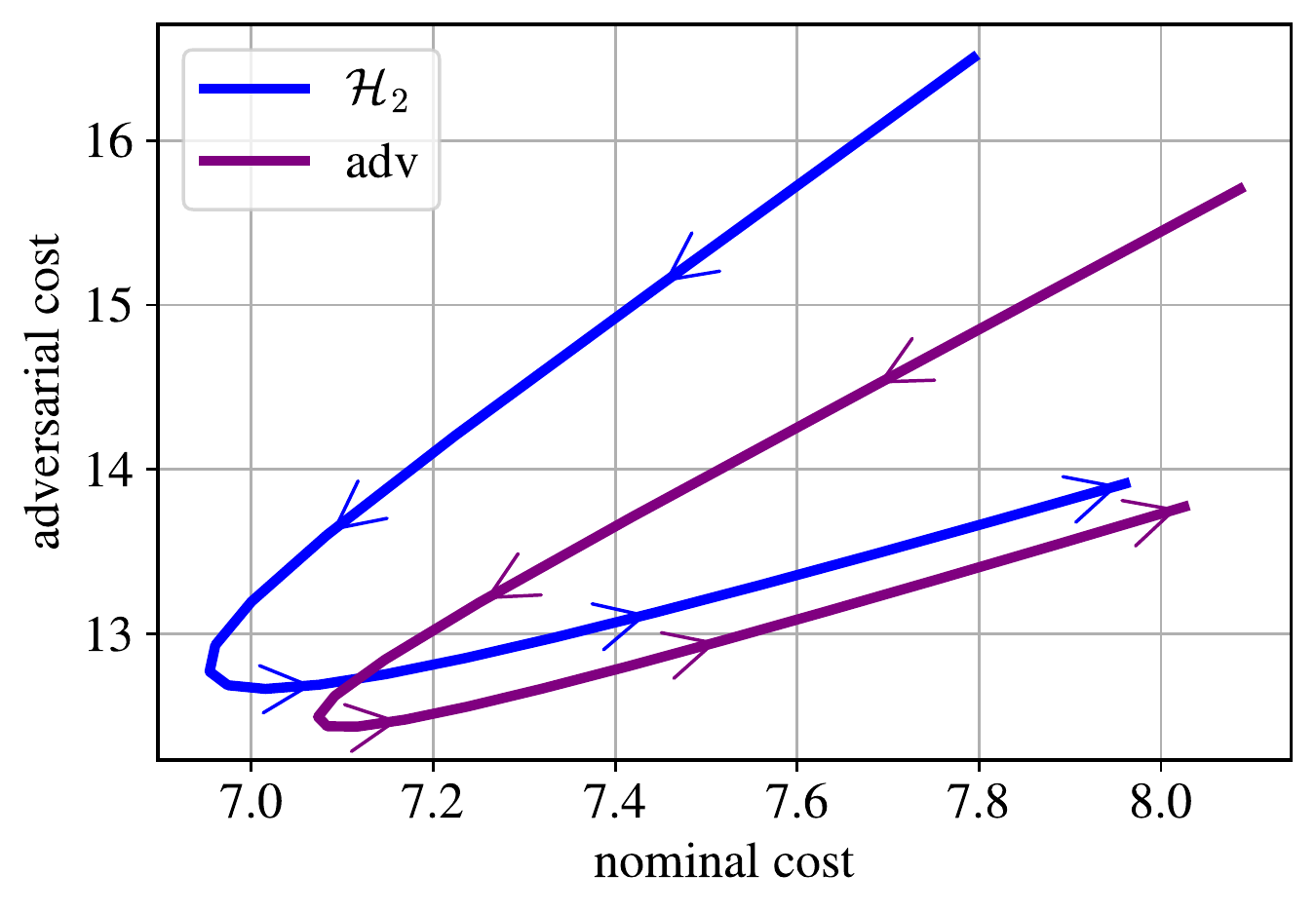}}
    \caption{ 
        (a) For each value of the system parameter $\rho$, the tradeoff curves are generated by synthesizing adversarially robust controllers for adversarial levels $\varepsilon$ ranging from $[0, 0.1]$, and then evaluating both their nominal cost, and their adversarial cost at level $\varepsilon=0.1$. (b) We plot the nominal vs.\ adversarial performance (at level $\varepsilon=0.1$) of the nominal LQR controller and the adversarially robust controller at level $\varepsilon=0.1$, as the system parameter $\rho$ ranges from $0.3$ to $1.2$. The value of $\rho$ increases in the directions of the arrows. 
     }
\end{figure}
} \else {
\begin{figure}[t]
    \label{fig:tradeoff curves}
    \centering
    \subfloat[Tradeoff curves]{\label{fig:tradeoff curves}\includegraphics[width=0.45\textwidth]{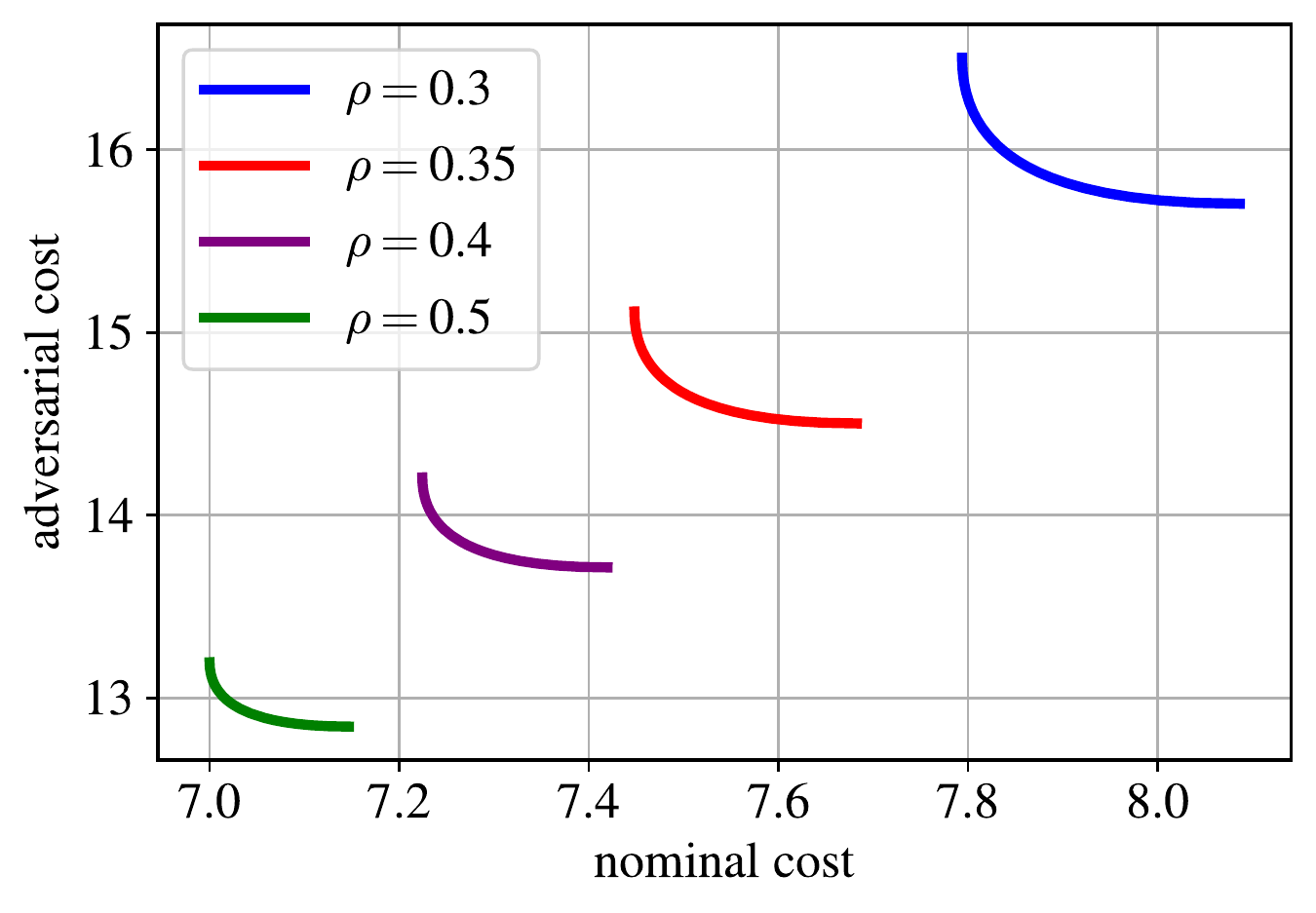}} 
    \subfloat[Tradeoff envelope]{\label{fig:tradeoff envelope}\includegraphics[width=0.45\textwidth]{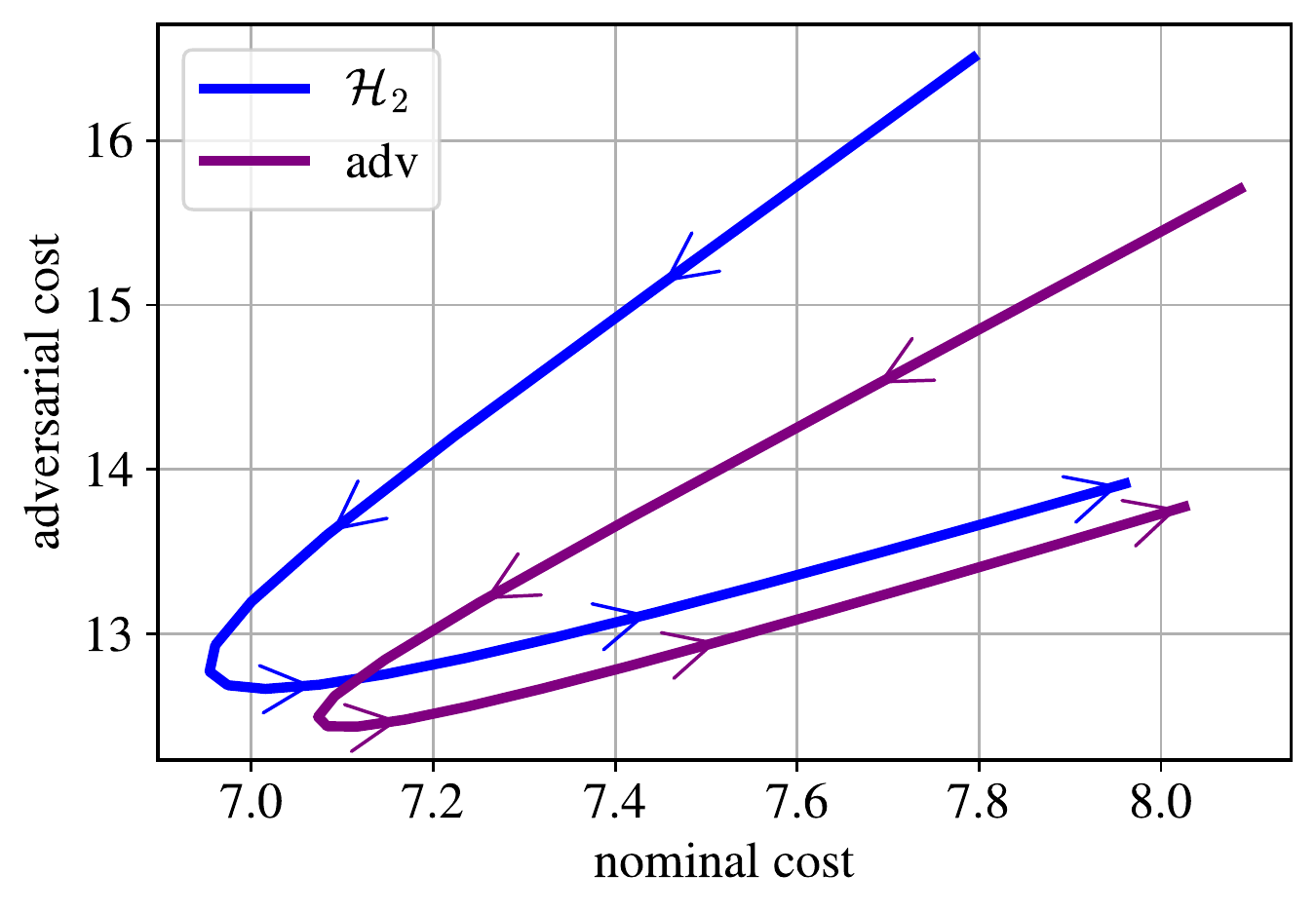}}
    \caption{ 
        (a) For each value of the system parameter $\rho$, the tradeoff curves are generated by synthesizing adversarially robust controllers for adversarial levels $\varepsilon$ ranging from $[0, 0.1]$, and then evaluating both their nominal cost, and their adversarial cost at level $\varepsilon=0.1$. (b) We plot the nominal vs.\ adversarial performance (at level $\varepsilon=0.1$) of the nominal LQR controller and the adversarially robust controller at level $\varepsilon=0.1$, as the system parameter $\rho$ ranges from $0.3$ to $1.2$. The value of $\rho$ increases in the directions of the arrows. 
     }
\end{figure}
} \fi

\ifnum\value{cdc}>0{
\begin{figure*}[t]
    \centering
    \subfloat[Gaussian noise]{\label{fig: gaussian}\includegraphics[width=0.315\textwidth]{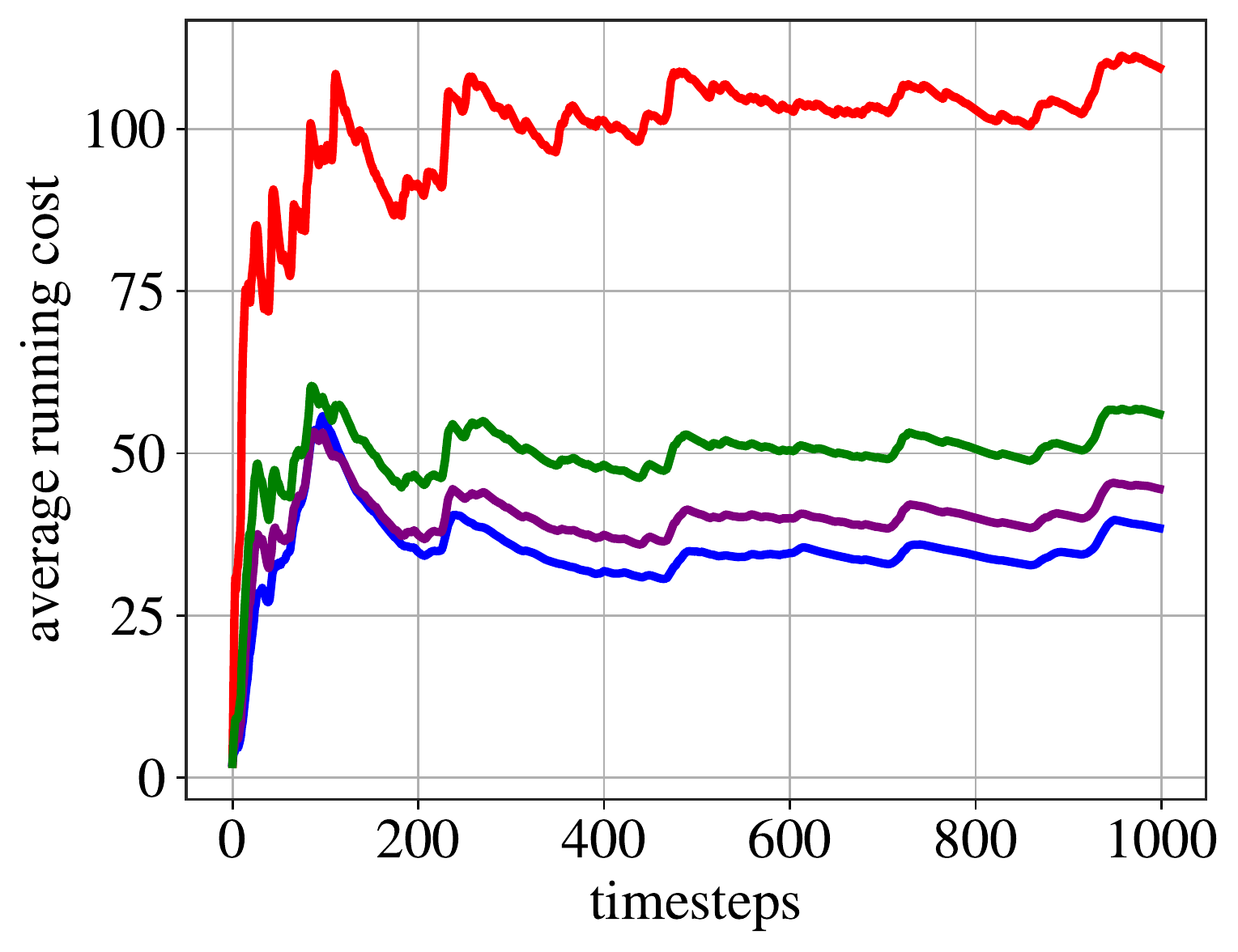}}
    \subfloat[Varying mean Gaussian]{\label{fig: varying mean gaussian}\includegraphics[width=0.3\textwidth]{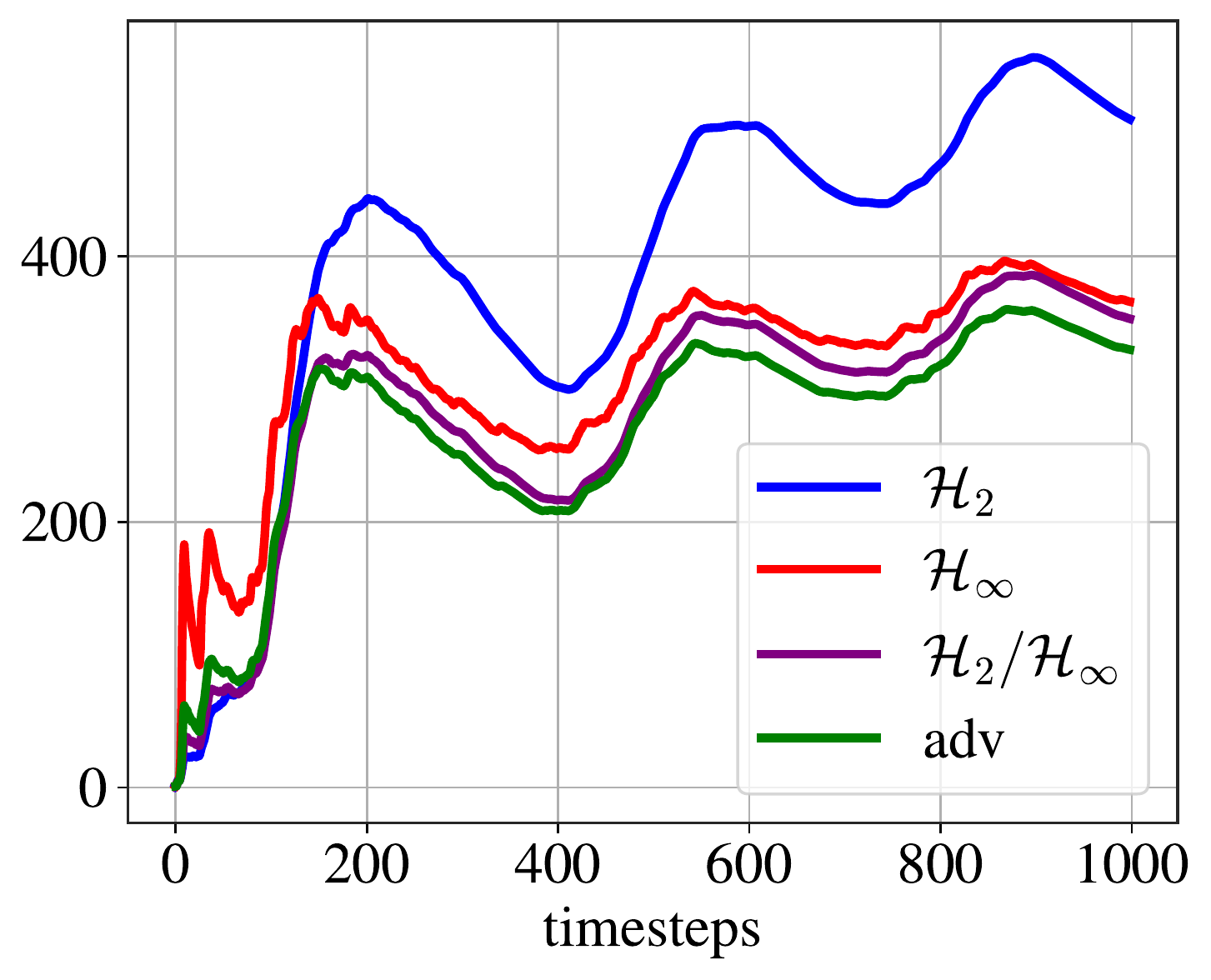}}
    \subfloat[Adversarial setting]{\label{fig: adversarial} \includegraphics[width=0.315\textwidth]{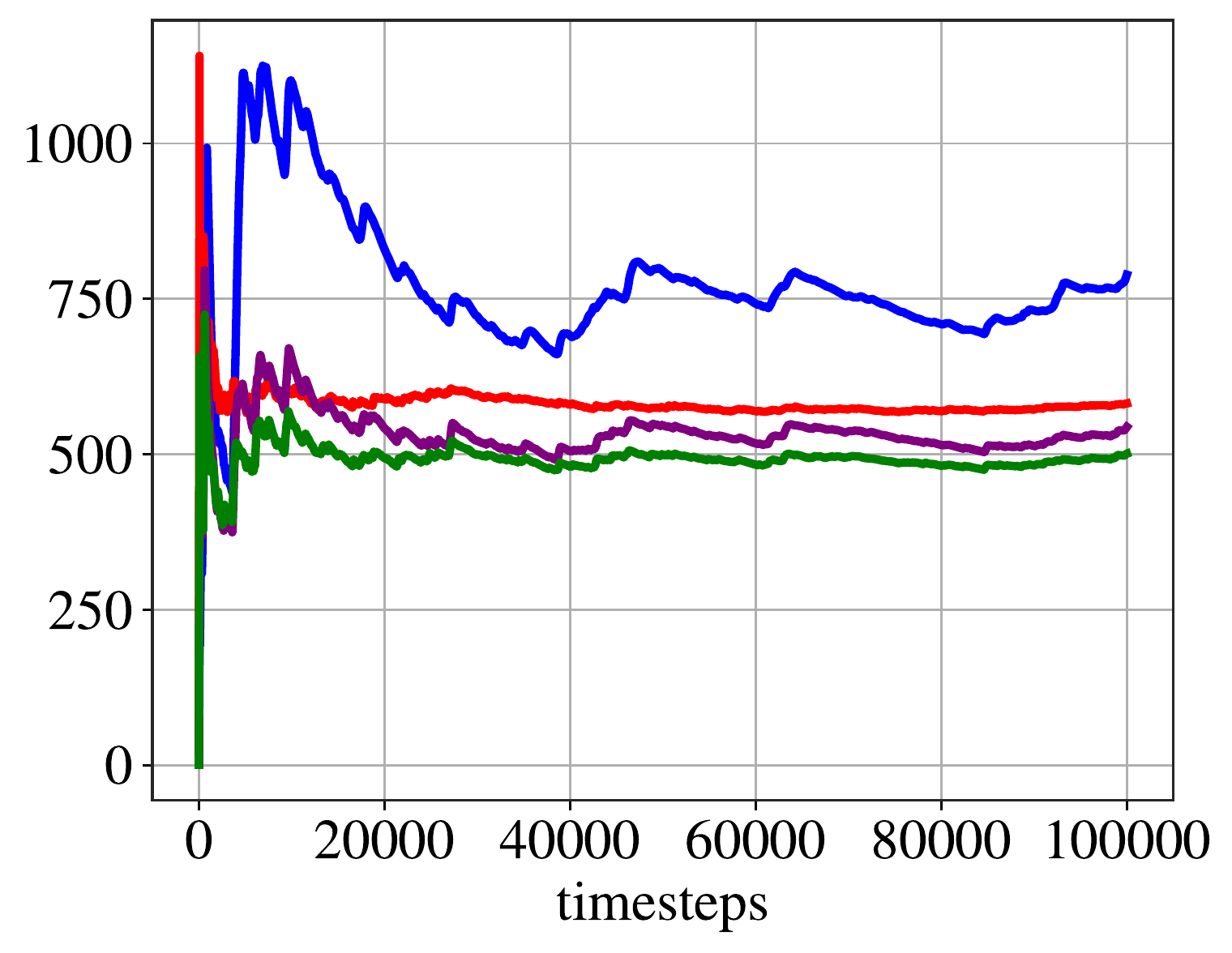}}
    \caption{
    Simulations of the performance of $\calH_2$, $\calH_\infty$, mixed $\calH_2/\calH_\infty$, and adversarially robust controllers on a linearized Boeing longitudinal flight control task. The average running cost at time $t$ is computed as $\frac{1}{t} \sum_{k=0}^t x_k^\top Qx_k + u_k^\top R u_k$.
    }
    \label{fig:my_label}
\end{figure*}
} \else {
\begin{figure}[t]
    \centering
    \subfloat[Gaussian noise]{\label{fig: gaussian}\includegraphics[width=0.315\textwidth]{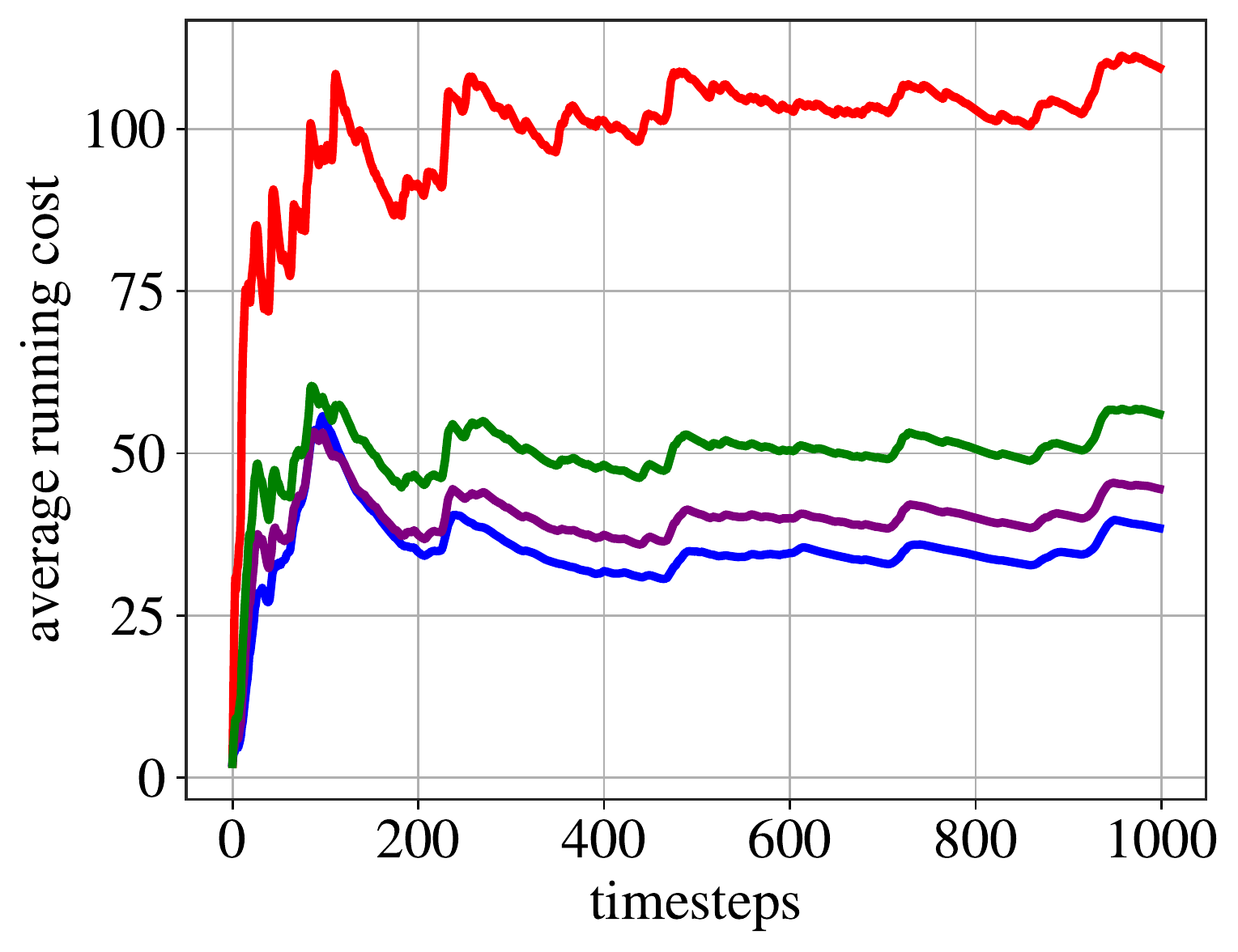}}
    \subfloat[Varying mean Gaussian]{\label{fig: varying mean gaussian}\includegraphics[width=0.3\textwidth]{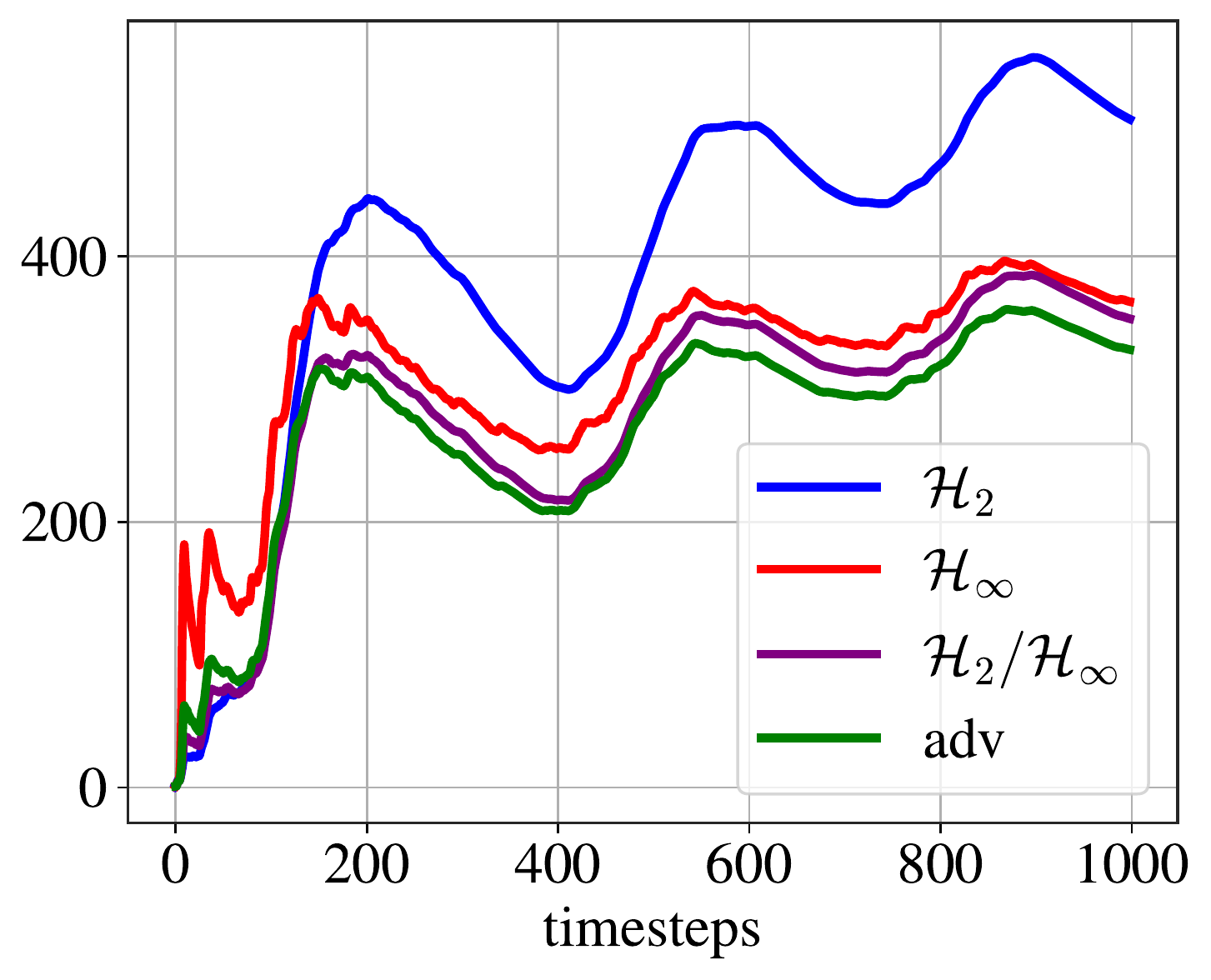}}
    \subfloat[Adversarial setting]{\label{fig: adversarial} \includegraphics[width=0.315\textwidth]{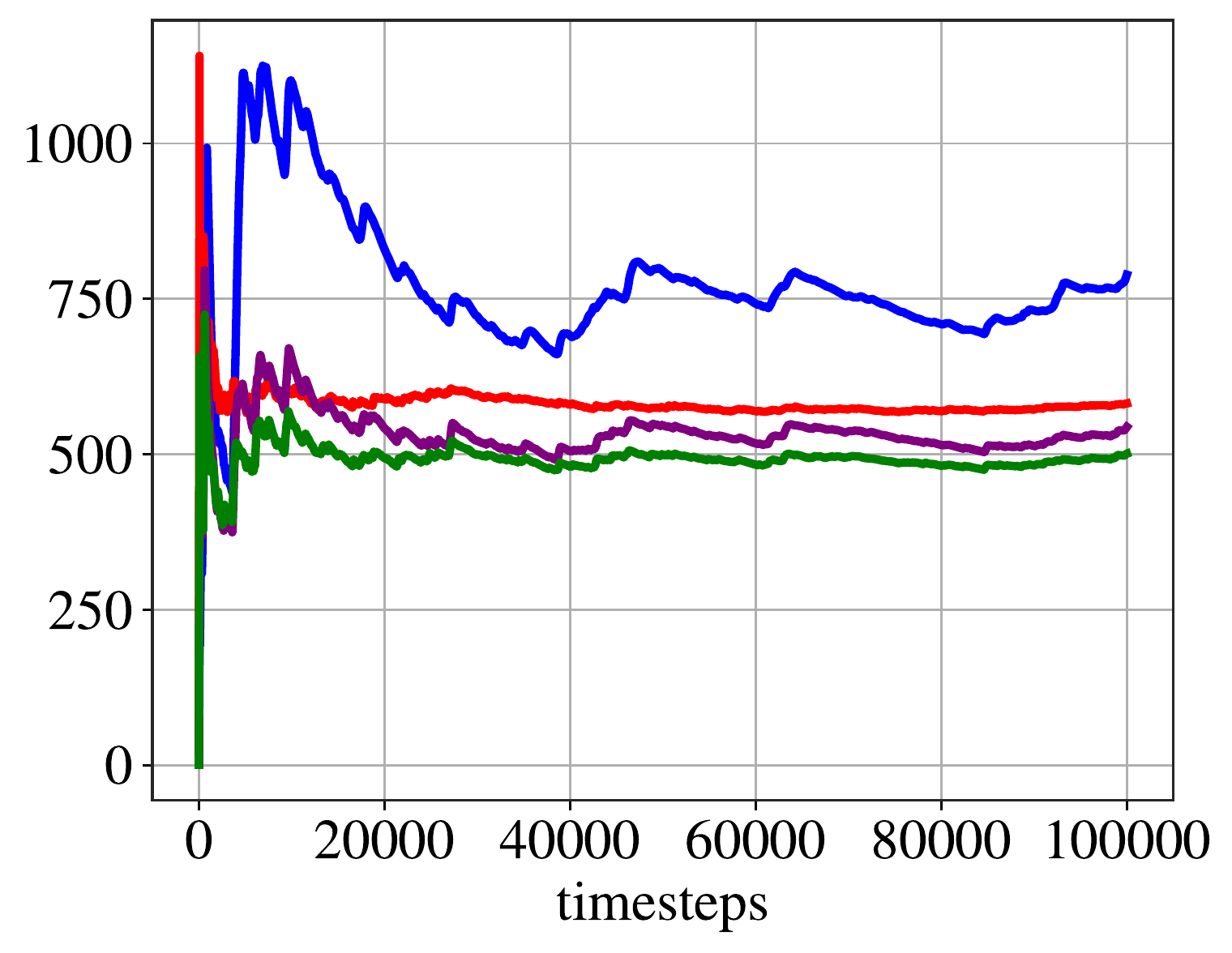}}
    \caption{
    Simulations of the performance of $\calH_2$, $\calH_\infty$, mixed $\calH_2/\calH_\infty$, and adversarially robust controllers on a linearized Boeing longitudinal flight control task. The average running cost at time $t$ is computed as $\frac{1}{t} \sum_{k=0}^t x_k^\top Qx_k + u_k^\top R u_k$.
    }
    \label{fig:my_label}
\end{figure}
} \fi

\paragraph{Impact of Controllability}
To illustrate the dependence of the tradeoff severity on system controllability, consider the simple integrator system defined by $(A, B, Q, R, \Sigma_w) = \paren{\bmat{1 & \rho \\ 0 & 1},\bmat{0 \\ 1},I, I, I}$
where we vary the controllability of the system by changing $\rho$. In particular, when $\rho$ is very small, the system has poor controllability, and as $\rho$ increases, controllability increases. In Figure~\ref{fig:tradeoff curves}, we consider the tradeoff curves traced out by evaluating the nominal cost and adversarial cost (evaluated with budget $\varepsilon = 0.1$) of adversarially robust controllers with adversarial budget varying between $\brac{0, 0.1}$, given a certain level of controllability determined by $\rho$. We observe that as controllability decreases, not only do the tradeoff curves shift upward, but the curves also widen. This corroborates the trend described in Theorem~\ref{thm: nominal cost gap upper bound}, where we show the bound on the nominal cost gap between adversarially robust and nominal controllers (i.e.\ width of the tradeoff curve) grows larger as controllability decreases. This trend is further illustrated in Figure~\ref{fig:tradeoff envelope}, where we plot the nominal and adversarial ($\varepsilon = 0.1$) costs attained by the $\calH_2$ and adversarially robust ($\varepsilon = 0.1$) controllers as a function of $\rho$, which we vary from $0.3$ to $1.2$. We observe that for small $\rho$, the system has poor controllability, and thus the distance between the $\calH_2$ and adversarially robust controller costs is large. As controllability increases, we see this gap decrease monotonically. We note that the costs do not monotonically improve as $\rho$ increases after some point, as the amplification of disturbances from the integrator will outstrip the benefits of improved controllability.

\paragraph{Performance of Adversarially Robust Control} We compare the performance of the adversarially robust LQR controller against $\calH_2$, $\calH_\infty$, and mixed $\calH_2/\calH_\infty$ control for the longitudinal flight control of the following linearized Boeing 747 system (see \cite{hong2021lecture} for further details) with
\ifnum\value{cdc}>0{
\begin{align*}
    x_{t+1} &= \bmat{0.99 & 0.03 & -.02 & -.32 \\ 0.01 & 0.47 & 4.7 & .00 \\ .02 & -.06 & .40 & .00 \\ 0.01 & -.04 & .72 & .99}x_t \\
    & \,\, +
    \bmat{.01 & .99 \\ -3.44 & 1.66\\ -.83 & .44 \\ -.47 & .25} u_t + w_t, \quad Q = I, R = I 
\end{align*}
} \else {
    \begin{align*}
    x_{t+1} &= \bmat{0.99 & 0.03 & -.02 & -.32 \\ 0.01 & 0.47 & 4.7 & .00 \\ .02 & -.06 & .40 & .00 \\ 0.01 & -.04 & .72 & .99}x_t 
    \bmat{.01 & .99 \\ -3.44 & 1.66\\ -.83 & .44 \\ -.47 & .25} u_t + w_t, \quad Q = I, R = I 
    \end{align*}
} \fi

under the presence of various disturbances $w_t$:
$w_t \overset{\mathrm{i.i.d.}}{\sim} \calN(0, I)$, $w_t \sim \calN(\sin(0.01 t), I)$, and $w_t$ given by zero mean, identity covariance Gaussian noise plus the worst case adversarial perturbation with power bounded by $\epsilon=0.5$. The initial condition is set to zero in all cases. 

The optimal adversarially robust controller is generated by running Algorithm~\ref{alg: AdvLQR} with $\varepsilon = 0.5$, $Q=I,R = I$, $\Sigma_w=I$ and the mixed $\calH_2/\calH_\infty$ controller was generated by fixing a $\calH_\infty$ norm bound of $\gamma = 1000$, and approximately minimizing the $\calH_2$ norm via the approach outlined in \cite{deoliveira1999lmi}. The $\calH_\infty$ norm bound $\gamma$ for mixed $\calH_2/\calH_\infty$ was chosen to achieve performance similar to the adversarially robust controller in the adversarial setting. 
We observe that in Figure~\ref{fig: gaussian} under the pure zero-mean Gaussian setting, the $\calH_2$ controller performs best.
We note that the adversarially robust controller has performance comparable to $\calH_2$.
Interestingly, the adversarially robust controller outperforms all three controllers on the varying-mean Gaussian disturbance in Figure~\ref{fig: varying mean gaussian}, which we note is an instance of a disturbance composed of both zero-mean stochastic and deterministic components. 
Lastly, as expected, the adversarially robust controller performs best in the adversarial setting of Figure~\ref{fig: adversarial}. 
\section{Conclusion}

We proposed an adversarially robust LQ control problem, and demonstrated that the optimal solution to this problem is given by a central static suboptimal $\calH_\infty$ controller. An interesting aspect of this solution is that unlike pure $\calH_2$ controllers, the adversarially robust controller depends upon the noise statistics. 
Experiments show that the adversarially robust controller performs similar to mixed $\calH_2/\calH_\infty$ controllers on a simple system. 

We used the adversarially robust control problem as a means to study performance-robustness tradeoffs in control. In particular, we derived quantitative upper and lower bounds on the performance gap between the nominal controller the adversarially robust controller. The bounds show that systems with uniformly good controllability will have small performance-robustness tradeoffs, while those with a highly controllable mode in the closed-loop system (viewing disturbances as inputs) will have a large performance-robustness tradeoff. These trends are corroborated by experiments with a simple linear system by tracing out tradeoff curves.  Directions for future work include the extension of the problem setting considered here to the measurement feedback setting, and considering how other adversarial training techniques can be translated to robust controller synthesis.

\section*{Acknowledgements}
Bruce D.\ Lee is supported by the Department of Defense through the National Defense Science \& Engineering Graduate Fellowship Program. The research of Hamed Hassani is supported by NSF Grants 1837253, 1943064, 1934876, AFOSR Grant FA9550-20-1-0111, and DCIST-CRA. Nikolai Matni is funded by NSF awards CPS-2038873, CAREER award ECCS-2045834, and a Google Research Scholar award. 

\bibliographystyle{abbrvnat}

\bibliography{refs}
\appendix

\section{Proofs from Section \ref{sec: advrobust lq control}}

\subsection{Proof of Lemma~\ref{lem:adversarial LQR sol}}
\textit{Proof:}
First define the cost to go for a horizon $T$ at step $\tau$. 
\[
    V_{\tau, T}(x_\tau) := \min_{u_{\tau:T-1}} \Ex_{w_{\tau:T-1}} \brac{ \max_{\substack{\delta_{\tau:T-1} \\ causal}} x_T^\top Q x_T + \sum_{t=\tau}^{T-1} x_t Q x_t + u_t^\top R u_t - \gamma^2 \delta_t^\top \delta_t}.
\]
Note that for any $T$, $V_{T,T}(x) = x^\top Q x$. We will show inductively that $V_{t,T}(x) = x^\top P_{t,T} x + q_{t,T}$ for any $t$. To do so, begin with the inductive hypothesis $V_{t+1, T}(x) = x^\top P_{t+1, T} x + q_{t+1}$. Applying this hypothesis, we have
\begin{align*}
     V_{\tau, T}(x_\tau) &= x_\tau^\top Q x_\tau + \min_{u_\tau} \curly{u_\tau^\top R u_\tau + \min_{u_{\tau+1:T-1}}  \Ex_{w_{\tau:T-1}} \brac{ \max_{\substack{\delta_{\tau:T-1} \\ causal}} x_T^\top Q x_T -\delta_\tau^\top \delta_\tau + \sum_{t=\tau+1}^{T-1} x_t Q x_t + u_t^\top R u_t - \gamma^2 \delta_t^\top \delta_t}} \\
    &= x_\tau^\top Q x_\tau + \min_{u_\tau} \bigg\{u_\tau^\top R u_\tau \\
    &+ \min_{u_{\tau+1:T-1}}  \Ex_{w_\tau} \brac{\max_{\delta_\tau} -\gamma^2 \delta_\tau^\top \delta_\tau + \Ex_{w_{\tau+1:T-1}} \brac{\max_{\substack{ \delta_{\tau+1:T-1} \\ causal}} x_T^\top Q x_T  + \sum_{t=\tau+1}^{T-1} x_t Q x_t + u_t^\top R u_t - \gamma^2 \delta_t^\top \delta_t}} \bigg\}  \\
     &= x_\tau^\top Q x_\tau + \min_{u_\tau} \curly{u_\tau^\top R u_\tau + \min_{u_{\tau+1:T-1}}  \Ex_{w_\tau} \brac{\max_{\delta_\tau} -\gamma^2 \delta_\tau^\top \delta_\tau + V_{\tau+1,T}(Ax_\tau + Bu_\tau + w_\tau + \delta_\tau)}} \\
     &= x_\tau^\top Q x_\tau + \min_{u_\tau} \bigg\{u_\tau^\top R u_\tau \\
     &+ \Ex_{w_\tau} \brac{\max_{\delta_\tau} -\gamma^2 \delta_\tau^\top \delta_\tau + (Ax_\tau + Bu_\tau + w_\tau + \delta_\tau)^\top P_{\tau+1, T} (Ax_\tau + Bu_\tau + w_\tau + \delta_\tau) + q_{\tau+1, T}}\bigg\}
\end{align*}
Letting $\phi_t := Ax_t + Bu_t + w_t$, we see that the maximization with respect to $\delta_\tau$ is equivalent to
\begin{align*}
    \max_{\delta_\tau} \delta_\tau^\top \paren{P_{\tau+1, T} - \gamma^2 I} \delta_\tau^\top + 2 \phi_\tau^\top P_{\tau+1, T} \delta_\tau.
\end{align*}
If $P_{\tau+1, T} \prec \gamma^2 I$, then there is a unique optimal solution given by $\delta_\tau = -(P_{\tau+1, T} - \gamma^2 I)^{-1} P_{\tau+1, T} \phi_\tau$. Substituting this back in, we see that 
\begin{align*}
    V_{\tau, T}(x_\tau) &= q_{\tau+1, T} + x_\tau^\top Q x_\tau + \min_{u_\tau} \bigg\{u_\tau^\top R u_\tau + \Ex_{w_\tau} \bigg \lbrack \phi_\tau^\top (-\gamma^2 P_{\tau+1, T}(P_{\tau+1, T} - \gamma^2 I)^{-2} P_{\tau+1, T} \\
    &\quad \quad + (I - P_{\tau+1, T}(P_{\tau+1, T} - \gamma^2 I)^{-1}) P_{\tau+1, T} (I - (P_{\tau+1, T} - \gamma^2 I)^{-1} P_{\tau+1, T})) \phi_\tau \bigg \rbrack\bigg\} \\
    &= q_{\tau+1, T} + x_\tau^\top Q x_\tau + \min_{u_\tau} \bigg\{u_\tau^\top R u_\tau + \Ex_{w_\tau} \bigg \lbrack \phi_\tau^\top (P_{\tau+1, T} + P_{\tau+1, T}(\gamma^2 I - P_{\tau+1, T})^{-1} P_{\tau+1}) \phi_\tau \bigg \rbrack\bigg\} \\
\end{align*}
Defining $M_{\tau+1, T} := P_{\tau+1, T} + P_{\tau+1, T}(\gamma^2 I - P_{\tau+1, T})^{-1} P_{\tau+1, T}$, the above expression reduces to
\begin{align*}
     q_{\tau+1, T} &+ x_\tau^\top Q x_\tau + \min_{u_\tau} \bigg\{u_\tau^\top R u_\tau + \Ex_{w_\tau} \bigg \lbrack \phi_\tau^\top M_{\tau+1} \phi_\tau \bigg \rbrack\bigg\} \\
     &=  q_{\tau+1, T} + x_\tau^\top Q x_\tau + \min_{u_\tau} \bigg\{u_\tau^\top R u_\tau + (Ax_t + Bu_t)^\top M_{\tau+1,T} (Ax_t + Bu_t)\bigg\} + \trace(M_{\tau+1,T} \Sigma_w)
\end{align*}
Minimizing with respect to $u_\tau$ provides
\[
    u_\tau = -(R + B^\top M_{\tau+1, T} B)^{-1} B^\top M_{\tau+1, T} A x_\tau
\]
Substituting this back in, we have that
\begin{align*}
    V_{\tau, T}(x_\tau) &= q_{\tau+1, T} + \trace(M_{\tau+1, T} \Sigma_w)  + x_\tau^\top \paren{ Q + A^\top M_{\tau+1, T} A - A^\top M_{\tau+1, T} B(R+B^\top M_{\tau+1 ,T}B)^{-1} B^\top 
    M_{\tau+1, T} A} x_\tau  \\
    &=: q_{\tau, T} + x_\tau^\top P_{\tau, T}x_\tau
\end{align*}

Thus we see that the cost to go is given as 
\begin{align*}
    V_{\tau, T}(x_\tau) &= q_{\tau, T} + x_\tau^\top P_{\tau, T} x_\tau 
\end{align*}
where 
\begin{align*}
    P_{\tau, T} &= Q + A^\top M_{\tau+1, T} A - A^\top M_{\tau+1, T} B(R+B^\top M_{\tau+1 ,T}B)^{-1} B^\top 
    M_{\tau+1, T} A \\
     M_{\tau, T} &= P_{\tau, T} +  P_{\tau, T}\paren{\gamma^2 I - P_{\tau, T}}^{-1} P_{\tau, T} \\
    q_{\tau, T} &= q_{\tau+1, T+1} +  \trace(M_{\tau+1, T} \Sigma_w) \\
    q_{T,T} &= 0\\ 
    P_{T,T} &= Q
\end{align*}
Taking the limit as $T \to \infty$, and denoting $P_\tau := P_{\tau, \infty}$ we see that if the iteration
\begin{align*}
    P_{\tau} &= Q + A^\top M_{\tau+1} A - A^\top M_{\tau+1} B(R+B^\top M_{\tau+1}B)^{-1} B^\top 
    M_{\tau+1} A \\
     M_{\tau} &= P_{\tau} +  P_{\tau}\paren{\gamma^2 I - P_{\tau}}^{-1} P_{\tau} 
\end{align*}
converges to a fixed point $P$, then the fixed point must satisfy 
\begin{align*}
    P &= Q + A^\top M A - A^\top M B(R+B^\top M B)^{-1} B^\top 
    M A \\
    M &= P +  P \paren{\gamma^2 I - P^{-1}} P.
\end{align*}
We see that finding a pair $(P,M)$ that satisfy the above conditions is equivalent to finding a $P$ such that 
\begin{align}\label{eq: gamma-suboptimal DARE}
    P = Q + A^\top P A - A^\top P \bmat{B & I}\paren{\bmat{B & I}^\top P \bmat{B & I} + \bmat{R & \\ & -\gamma^2 I}}^{-1}\bmat{B & I}^\top P A
\end{align}
To see that this is the case, note that
\begin{align*}
    &P + P^\top \bmat{B & I} \bmat{B^\top P B + R & B^\top P \\ P B & P - \gamma^2 I}^{-1} \bmat{B & I}^\top P   \\
    & \hspace{-.0 in}= P + P\bmat{B^\top \\ I}^\top \bmat{ \Psi^{-1}  & \Psi^{-1}B^\top P(\gamma^2 I- P)^{-1}  \\ (\gamma^2 I-P)^{-1} P B \Psi^{-1} & (P - \gamma^2 I)^{-1} + (\gamma^2 I-P)^{-1} PB \Psi^{-1} P(\gamma^2 I- P)^{-1}} \bmat{B^\top \\ I} P \\
    &= M + MB(B^\top MB^\top +R)^{-1}  B^\top M
\end{align*}
where $\Psi = B^\top M B  + R$. The first equality above follows from the inversion formula for a block $2\times2$ matrix. 

By our assumption that there exists a positive semidefinite solution to \eqref{eq:modified DARE}, the iterations converges to this quantity. This can be seen by verifying the monotonicty of the sequence $P_\tau$, as in Lemma 3.3 of \cite{basar1991hinf}. 

We thus have that in the limit, the cost converges to $\trace(M\Sigma_w)$,
where $M = P + P(\gamma^2 I - P)^{-1}P$ and $P$ solves
\begin{align*}
    P = Q + A^\top P A - A^\top P \bmat{B & I}\paren{\bmat{B & I}^\top P \bmat{B & I} + \bmat{R & \\ & -\gamma^2 I}}^{-1}\bmat{B & I}^\top P A.
\end{align*}
Thus a steady state adversarially robust LQR controller at level $\gamma$ is given by 
\begin{equation*}
\begin{aligned}
\label{eq:advrob controller}
    u_t &= K x_t \\
    K &= -(R + B^\top M B)^{-1} B^\top M A \\
    M &= P+P(\gamma^2 I - P)^{-1}P \\
    P &\mbox{ solves \eqref{eq:modified DARE}}
\end{aligned}
\end{equation*}
and the optimal adversarial perturbation is given by
\[
    \delta_t = (\gamma^2 I - P)^{-1} P ((A+BKx_t) + w_t),
\]
To verify that the controller is stabilizing, it suffices to note that the cost is finite. Leveraging the fact that $(Q^{1/2}, A)$ is detectable thus tells us that the state must converge to zero in the absence of noise. 

For necessity of the given conditions, first note that if there is no PSD solution to the Riccati equation, then the value does not converge. If the condition $P \prec \gamma^2 I$ is not satisfied, then the adversarial maximization problem is not strictly concave, and it may supply arbitrarily large perturbations. 
\hfill $\blacksquare$

\section{Proof of Theorem~\ref{thm:adversarial controller solution}}
\label{appendix: advlqr thm proof}

The following lemma will be useful. 

\begin{lemma} (Ergodicity)
\label{lem:ergodicity}

Let our controller be given as $u_t = K x_t$, and define
\[
    P = \dare(A+BK, I, Q, -\gamma^2 I)
\]
As long as 
\[
(\gamma^2 I - P)^{-1} P (A+BK)
\]
has eigenvalues in the open unit disc,  
\begin{align*}
    \limsup_{T \to \infty}
    \frac{1}{T} \Ex \brac{\max_{\delta \mbox{ causal}} x_T^\top Q x_T + \sum_{t=0}^{T-1} x_t Q x_t + u_t^\top R u_t - \gamma^2 \delta_t^\top \delta_t} \\
    = \limsup_{T \to \infty}
    \frac{1}{T} \max_{\delta \mbox{ causal}} x_T^\top Q x_T + \sum_{t=0}^{T-1} x_t Q x_t + u_t^\top R u_t - \gamma^2 \delta_t^\top \delta_t.
\end{align*}

\end{lemma}
\textit{Proof:}
To see that this is so, recall that the optimal adversarial perturbation was shown to be given by 
\[
    \delta_t = -(P - \gamma^2 I)^{-1} P \paren{(A+BK)x_t + w_t}
\]  
and under this perturbation, the dynamics become 
\[
    x_{t+1} = \brac{I - (P-  \gamma I)^{-1} P}\paren{(A+BK)x_t + w_t}
\]
Then $\bmat{x_t \\ w_t}$ is an ergodic process. In this case, we have that
\begin{align*}
    =&\limsup_{T \to \infty}
    \frac{1}{T} \max_{\delta \mbox{ causal}} x_T^\top Q x_T + \sum_{t=0}^{T-1} x_t Q x_t + u_t^\top R u_t - \gamma^2 \delta_t^\top \delta_t \\
    =&\limsup_{T \to \infty}
    \frac{1}{T} x_T^\top Q x_T + \sum_{t=0}^{T-1} x_t (Q+K^\top R K) x_t \\ 
    &\quad- \gamma^2 ((A+BK)x_t. + w_t)^\top \brac{I - (P-  \gamma I)^{-1} P}^\top \brac{I - (P-  \gamma I)^{-1} P} (A+BK)x_t + w_t)  \\
    = &\limsup_{t \to \infty} \Ex\bigg[x_t (Q+K^\top R K) x_t \\ &\quad - \gamma^2 ((A+BK)x_t. + w_t)^\top \brac{I - (P-  \gamma I)^{-1} P}^\top \brac{I - (P-  \gamma I)^{-1} P} (A+BK)x_t + w_t)\bigg] \\
    = &\limsup_{T \to \infty}
    \frac{1}{T} \Ex\bigg[x_T^\top Q x_T + \sum_{t=0}^{T-1} x_t (Q+K^\top R K) x_t \\ &\quad - \gamma^2 ((A+BK)x_t. + w_t)^\top \brac{I - (P-  \gamma I)^{-1} P}^\top \brac{I - (P-  \gamma I)^{-1} P} (A+BK)x_t + w_t)\bigg] \\
    =&\limsup_{T \to \infty}
    \frac{1}{T} \Ex \brac{\max_{\delta \mbox{ causal}} x_T^\top Q x_T + \sum_{t=0}^{T-1} x_t Q x_t + u_t^\top R u_t - \gamma^2 \delta_t^\top \delta_t}
\end{align*}
\hfill $\blacksquare$

Now we are ready to prove Theorem~\ref{thm:adversarial controller solution}. 

\textit{Proof:}
First note that optimization problem
\[
     \max_{\substack{\mathbf{\delta} \mbox{ causal} \\ \norm{\mathbf{\delta}}_{\ell_2}^2 \leq T \varepsilon}} x_T^\top Q x_T + \sum_{t=0}^{T-1} x_t^\top Q x_t + u_t^\top R u_t
\]
is the maximization of a quadratic function subject to a quadratic constraint, and therefore satisfies strong duality (see \cite{boyd2004convex}, Appendix B.1). Therefore, maximizing subject to the quadratic constraint is equivalent to maximizing the Lagrangian, then minimizing over the dual variable. 
\begin{align*}
    \max_{\substack{\mathbf{\delta} \mbox{ causal} \\ \norm{\mathbf{\delta}}_{\ell_2}^2 \leq T \varepsilon}} x_T^\top Q x_T + \sum_{t=0}^{T-1} x_t^\top Q x_t + u_t^\top R u_t = \min_\gamma 
    \gamma^2 \varepsilon T + \max_{\mathbf{\delta} \mbox{ causal}} x_T^\top Q x_T + \sum_{t=0}^{T-1} x_t^\top Q x_t + u_t^\top R u_t - \gamma^2 \delta_t^\top \delta_t.
\end{align*}
Then \eqref{eq:hardadversarialLQRobjective} is equivalent to 
\begin{align*}
    \minimize_{u} \limsup_T \frac{1}{T} \Ex \brac{\min_\gamma  \gamma^2 \varepsilon T + \max_{\mathbf{\delta} \mbox{ causal}} x_T^\top Q x_T + \sum_{t=0}^{T-1} x_t^\top Q x_t + u_t^\top R u_t - \gamma^2 \delta_t^\top \delta_t}.
\end{align*}
Our next step is to show that we may pull the minimization over $\gamma$ outside of the expectation. To see that this is the case, we will make use of ergodicity. 

By the assumption that $(A,B)$ is stabilizable, and $(Q^{1/2}, A)$ is detectable, we have that 
 \begin{align*}
    \infty &> \min_\gamma \gamma^2 \varepsilon T + \limsup_{T \to \infty} \min_{u_0, \dots, u_{T-1}}
    \frac{1}{T} \Ex \brac{\max_{\delta \mbox{ causal}} x_T^\top Q x_T + \sum_{t=0}^{T-1} x_t Q x_t + u_t^\top R u_t - \gamma^2 \delta_t^\top \delta_t} \\
    &= \min_\gamma \gamma^2 \varepsilon T + \liminf_{T \to \infty} \min_{u_0, \dots, u_{T-1}}
    \frac{1}{T} \Ex \brac{\max_{\delta \mbox{ causal}} x_T^\top Q x_T + \sum_{t=0}^{T-1} x_t Q x_t + u_t^\top R u_t - \gamma^2 \delta_t^\top \delta_t}.
\end{align*}
Now, for each $\gamma$ let $K_\gamma$ be the feedback controller defined in Lemma~\ref{lem:adversarial LQR sol}, and let $u_t = K_\gamma x_t$. Then
\begin{align*}
    & \quad \quad \, \, \,\min_\gamma \gamma^2 \varepsilon T + \limsup_{T \to \infty}  
    \frac{1}{T} \Ex \brac{\max_{\delta \mbox{ causal}} x_T^\top Q x_T + \sum_{t=0}^{T-1} x_t Q x_t + u_t^\top R u_t - \gamma^2 \delta_t^\top \delta_t} \\
    &=\Ex \brac{\min_\gamma \gamma^2 \varepsilon T + \limsup_{T \to \infty}
    \frac{1}{T} \Ex \brac{\max_{\delta \mbox{ causal}} x_T^\top Q x_T + \sum_{t=0}^{T-1} x_t Q x_t + u_t^\top R u_t - \gamma^2 \delta_t^\top \delta_t}} \\
    &= \Ex \brac{ \min_\gamma \gamma^2 \varepsilon T + \limsup_{T \to \infty} 
    \frac{1}{T} \max_{\delta \mbox{ causal}} x_T^\top Q x_T + \sum_{t=0}^{T-1} x_t Q x_t + u_t^\top R u_t - \gamma^2 \delta_t^\top \delta_t} \\
    &\geq \limsup_{T \to \infty} \frac{1}{T} \Ex \brac{ 
     \min_\gamma \gamma^2 \varepsilon T + \max_{\delta \mbox{ causal}} x_T^\top Q x_T + \sum_{t=0}^{T-1} x_t Q x_t + u_t^\top R u_t - \gamma^2 \delta_t^\top \delta_t}.
\end{align*}
Where the first equality is permitted by the fact that first line is not random. The second line follow from ergodicity (Lemma \ref{lem:ergodicity}). The inequality is a result of Fatou's Lemma and the $\min \max$ inequality. We can similarly obtain an upper bound:
\begin{align*}
    & \quad \quad \, \, \,\min_\gamma \gamma^2 \varepsilon T + \liminf_{T \to \infty}  
    \frac{1}{T} \Ex \brac{\max_{\delta \mbox{ causal}} x_T^\top Q x_T + \sum_{t=0}^{T-1} x_t Q x_t + u_t^\top R u_t - \gamma^2 \delta_t^\top \delta_t} \\
    &=\Ex \brac{\min_\gamma \gamma^2 \varepsilon T + \liminf_{T \to \infty}
    \frac{1}{T} \Ex \brac{\max_{\delta \mbox{ causal}} x_T^\top Q x_T + \sum_{t=0}^{T-1} x_t Q x_t + u_t^\top R u_t - \gamma^2 \delta_t^\top \delta_t}} \\
    &= \Ex \brac{ \min_\gamma \gamma^2 \varepsilon T + \liminf_{T \to \infty} 
    \frac{1}{T} \max_{\delta \mbox{ causal}} x_T^\top Q x_T + \sum_{t=0}^{T-1} x_t Q x_t + u_t^\top R u_t - \gamma^2 \delta_t^\top \delta_t} \\
    &\leq \liminf_{T \to \infty} \frac{1}{T} \Ex \brac{ 
     \min_\gamma \gamma^2 \varepsilon T + \max_{\delta \mbox{ causal}} x_T^\top Q x_T + \sum_{t=0}^{T-1} x_t Q x_t + u_t^\top R u_t - \gamma^2 \delta_t^\top \delta_t}.
\end{align*}
Combining these results, we see that
\begin{align*}
    \limsup_{T \to \infty} \frac{1}{T} \Ex \brac{ 
     \min_\gamma \gamma^2 \varepsilon T + \max_{\delta \mbox{ causal}} x_T^\top Q x_T + \sum_{t=0}^{T-1} x_t Q x_t + u_t^\top R u_t - \gamma^2 \delta_t^\top \delta_t} \\
     = \min_\gamma \gamma^2 \varepsilon + \limsup_{T \to \infty}  
    \frac{1}{T} \Ex \brac{\max_{\delta \mbox{ causal}} x_T^\top Q x_T + \sum_{t=0}^{T-1} x_t Q x_t + u_t^\top R u_t - \gamma^2 \delta_t^\top \delta_t}.
\end{align*}
Therefore, we see that the optimal solution is given by Lemma~\ref{lem:adversarial LQR sol} for some properly chosen $\gamma$. If $\gamma$ becomes smaller than $\gamma_\infty$, the problem will be unbounded. Now, observe that i) the power of the optimal adversary for the problem at some level $\gamma$, $\lim_t \Ex\brac{\paren{\delta_t^{\gamma}}^\top \delta_t^\gamma}$ is decreasing as $\gamma$ increases, and ii) the power of the optimal adversary for the hard constrained problem \eqref{eq:adversarialLQRobjective} will equal $\varepsilon$. Then bisecting such that the adversarial power of the solution to Lemma~\ref{lem:adversarial LQR sol} provides the proper level $\gamma$. 

\hfill $\blacksquare$ 

\section{Proofs for Section~\ref{s: performance robustness tradeoffs}}

We introduce the following recurring fact: given a solution $P_\gamma$ to the DARE in Lemma~\ref{lem:adversarial LQR sol}, we have
\begin{align} \label{eq: x0'Px0}
    x_0^\top P_\gamma x_0 &= \limsup_{T \to \infty}  
    \frac{1}{T} \paren{\max_{\delta \mbox{ causal}} x_T^\top Q x_T + \sum_{t=0}^{T-1} x_t \paren{Q + K_{\gamma_1}^\top R K_{\gamma_1}} x_t - \gamma_1^2 \delta_t^\top \delta_t }.
\end{align}
In other words, $x_0^\top P_\gamma x_0$ is the soft-constrained cost of the optimal adversarially robust controller against the optimal adversary at level $\gamma$ in the noiseless setting, with initial condition $x_0$.

\subsection{Proof of Lemma~\ref{lem: controller gap to DARE gap}}
The following lemma from \cite{mania2019certainty} will be helpful. 
\begin{lemma}(Lemma 1 in \cite{mania2019certainty})
    \label{lem: upper bounding gap between optimizers}
    Let $f_1$ and $f_2$ be $\mu$-strongly convex twice differentiable functions functions. Let $x_1 = \argmin_x f_1(x)$ and $x_2 = \argmin_x f_2(x)$. Suppose $\norm{\nabla f_1(x_2)} \leq \varepsilon$. Then $\norm{x_1 - x_2} \leq \frac{\varepsilon}{\mu}$. 
\end{lemma}

We also consider an analogous lemma which provides a lower bound.
\begin{lemma}
    \label{lem: lower bounding gap between optimizers}
    Suppose $f_1$ and $f_2$ are $L$-Lipschitz function, and let $x_1$ minimize $f_1$ and $x_2$ minimize $f_2$. If $\norm{\nabla f_1(x_2)} \geq \varepsilon$, then $\norm{x_1 - x_2} \geq \frac{\varepsilon}{L}$.
\end{lemma}

\textit{Proof:}
By a Taylor expansion of $\nabla f_1$ about $x_1$, we see that
\begin{align*}
    \nabla f_1(x_2) = \nabla f_1 (x_1) + \nabla^2 f_1(\tilde x) (x_2 - x_1)
\end{align*}
where $\tilde x = t x_1 + (1-t) x_2$ for some $t \in [0,1]$. As $x_1$ minimizes $f_1$, we must have $\nabla f_1(x_1) = 0$. Futhermore, by the fact that $f_1$ is $L$ Lipschitz, we have that
\begin{align*}
    \norm{\nabla f_1(x_2)} = \norm{\nabla f_1 (x_1) + \nabla^2 f_1(\tilde x) (x_2 - x_1)} = \norm{\nabla^2 f_1(\tilde x) (x_2 - x_1)} \leq L \norm{x_2 - x_1}
\end{align*}
Dividing both sides of the above inequality by $L$ and leveraging the assumption that $\norm{\nabla f_1(x_2)}\leq \varepsilon$ provides the desired result. \hfill $\blacksquare$

Armed with these lemmas, we may prove Lemma~\ref{lem: controller gap to DARE gap}.

\textit{Proof:}
First, note that both functions are strongly convex with parameter $\sigma_{\min}(R + B^\top P B)$, and Lipschitz continuous with parameter $\norm{R+B^\top M B}$. 

For any $x$, by the fact that $\nabla f_2(u_2, x) = 0$, we have \begin{align*}
    \nabla f_1(u_2, x) &= \nabla f_1(u_2, x) - \nabla f_2(u_2, x) \\
    &= (R + B^\top M B) u_2 + B^\top M Ax - (R+B^\top P B) u_2 - B^\top P A x \\
    &= B^\top (M-P) (A+B K_2) x,
\end{align*}
and thus
\begin{align*}
    \norm{\nabla f_1(u_2, x)} = \norm{B^\top (M-P) (A+B K_2) x}.
\end{align*}
Then by application of Lemma~\ref{lem: upper bounding gap between optimizers} and Lemma~\ref{lem: lower bounding gap between optimizers}, we have that for any $x$, $\norm{u_1 - u_2}=\norm{(K_1- K_2)x}$ satisfies
\begin{align*}
    \frac{
    \norm{B^\top (M-P) (A+B K_2) x}}{\norm{R+B^\top M B}} \leq \norm{(K_1-K_2)x} \leq \frac{\norm{B^\top (M-P) (A+B K_2) x}}{\sigma_{\min}(R + B^\top P B)}.
\end{align*}
Taking the supremum over $\norm{x}=1$, we find
\begin{align*}
    \frac{
    \norm{B^\top (M-P) (A+B K_2)}}{\norm{R+B^\top M B}} \leq \norm{K_1-K_2} \leq \frac{\norm{B^\top (M-P) (A+B K_2) }}{\sigma_{\min}(R + B^\top P B)}.
\end{align*}

\hfill $\blacksquare$

\subsection{Proofs for Section~\ref{sec: upper bound}}\label{appendix: upper bound}

From the upper bound in Equation~\eqref{eq:cost gap upper bound}, we need to bound $\norm{\Sigma(K_\gamma)}$ and $\norm{K_\gamma - K_\star}$. For $\norm{\Sigma(K_\gamma)}$, we have by definition:
\begin{align}
    \norm{\Sigma(K_\gamma)} &:= \norm{\sum_{t=0}^\infty (A+BK_\gamma)^t \Sigma_w {(A+BK_\gamma)^\top}^t} \nonumber\\
    &= \sigma_w^2 \norm{\sum_{t=0}^\infty (A+BK_\gamma)^t {(A+BK_\gamma)^\top}^t} \nonumber\\
    &= \sigma_w^2 \norm{W_\infty(A+BK_\gamma, I)}, \label{eq: upper bound on Sigma(K_gamma)}
\end{align}
where we recall $W_\infty(A,B)$ is the controllability gramian, and we are guaranteed from Lemma~\ref{lem:adversarial LQR sol} that $\rho(A+BK_\gamma) < 1$, since $\gamma \geq \gamma_\infty$. We note that there exists a uniform constant upper bound on $\norm{W_\infty(A+BK_\gamma, I)}$ independent of $\gamma$, since $W_\infty(A+BK_{\gamma_\infty}, I)$ and $W_\infty(A+BK_\star, I)$ are both well-defined. From the intuition that $W_\infty(A+BK_\gamma, I)$ represents the ``disturbability'' of the closed-loop system $A+BK_\gamma$, i.e.\ the controllability of $A+BK_\gamma$ treating disturbances as inputs, one may conjecture some monotone behavior of $\norm{W_\infty(A + BK_\gamma, I)}$ between $\gamma = \gamma_\infty$ and $\gamma = \infty$ (i.e.\ going from the $\calH_\infty$ robust closed-loop system to the nominal closed-loop system), however proving such behavior of the curve drawn by $\gamma$ is outside the scope of this work. Alternatively, using norm bounds on the difference of block-Toeplitz matrices (such as those in \cite{mania2019certainty}), one can show that for sufficiently large $\gamma$, an upper bound on $\norm{W_\infty(A+BK_\gamma, I)}$ using only nominal system parameters can be derived, but these bounds tend to be overly conservative, as they throw out the structure provided by the adversarial LQ formulation. For illustrative purposes, we leave this characterization of $\norm{\Sigma(K_\gamma)}$ as is.

We may use Lemma~\ref{lem: controller gap to DARE gap} to get an upper bound on $\norm{K_\gamma - K_\star}$:
\begin{align}
    \norm{K_\gamma - K_\star} &\leq \frac{\norm{B^\top(M_\gamma - P_\star) (A+BK_\star)}}{\sigma_{\min}(R + B^\top P_\star B)}.
\end{align}
This reduces our problem to upper bounding $\norm{M_\gamma - P_\star}$ As discussed in Section~\ref{sec: upper bound}, this can be further upper bounded by
\begin{align*}
    \norm{M_\gamma - P_\star} &\leq \norm{P_\gamma - P_\star} + \frac{\norm{P_\gamma}^2}{\gamma^2 - \norm{P_\gamma}}.
\end{align*}
Therefore, our problem reduces to upper bounding $\norm{P_\gamma}$ and $\norm{P_\gamma - P_\star}$. For upper bounding $\norm{P_\gamma}$, we use the following lemma:
\begin{lemma}\label{lem: gamma ordering}
    Given $(A,B,Q^{1/2})$ stabilizable and detectable, for $\gamma_1 > \gamma_2 \geq \gamma_\infty$, we have
    \begin{align*}
        0 \prec P_{\gamma_1} \prec P_{\gamma_2} \preceq P_{\gamma_\infty} \preceq \gamma_\infty^2,
    \end{align*}
    where $P_{\gamma_1}$ and $P_{\gamma_2}$ are the solutions to the DARE in Lemma~\ref{lem:adversarial LQR sol} Equation~\eqref{eq:modified DARE} at levels $\gamma_1,\gamma_2$ respectively.
\end{lemma}

\textit{Proof:} this lemma follows by observing that since a smaller $\gamma$ corresponds to a smaller penalty on the adversary, for any initial condition, the corresponding adversarial LQ cost~\eqref{eq:adversarialLQRobjective} will be larger. To see this, let $\curly{\delta^{\gamma_1}_t}$ be optimal adversarial perturbations corresponding to level $\gamma_1$. Then $\curly{\delta^{\gamma_1}_t}$ are clearly feasible adversarial perturbations to the adversarial LQ instance at level $\gamma_2$, and since $\gamma_1 > \gamma_2$, the corresponding adversarial cost at level $\gamma_2$ must be larger. The maximal adversarial perturbations at level $\gamma_2$, $\curly{\delta^{\gamma_2}_t}$, can only incur higher cost than that, and thus we can show
\begin{align*}
    x_0^\top P_{\gamma_1}x_0 &= \limsup_{T \to \infty}  
    \frac{1}{T} \paren{\max_{\delta \mbox{ causal}} x_T^\top Q x_T + \sum_{t=0}^{T-1} x_t \paren{Q + K_{\gamma_1}^\top R K_{\gamma_1}} x_t - \gamma_1^2 \delta_t^\top \delta_t } \\
    &=: \limsup_{T \to \infty}  
    \frac{1}{T} \paren{x_T^\top Q x_T + \sum_{t=0}^{T-1} x_t \paren{Q + K_{\gamma_1}^\top R K_{\gamma_1}} x_t - \gamma_1^2 {\delta^{\gamma_1}_t}^\top \delta^{\gamma_1}_t } \\
    &< \limsup_{T \to \infty}  
    \frac{1}{T} \paren{x_T^\top Q x_T + \sum_{t=0}^{T-1} x_t \paren{Q + K_{\gamma_2}^\top R K_{\gamma_2}} x_t - \gamma_2^2 {\delta^{\gamma_1}_t}^\top \delta^{\gamma_1}_t } \\
    &\leq \limsup_{T \to \infty}  
    \frac{1}{T} \paren{\max_{\delta \mbox{ causal}} x_T^\top Q x_T + \sum_{t=0}^{T-1} x_t \paren{Q + K_{\gamma_2}^\top R K_{\gamma_2}} x_t - \gamma_2^2 \delta_t^\top \delta_t } \\
    &= x_0^\top P_{\gamma_2} x_0.
\end{align*}
Thus, we have shown $x_0^\top P_{\gamma_1} x_0 < x_0^\top P_{\gamma_2} x_0$ for any $x_0$, and thus $P_{\gamma_1} \prec P_{\gamma_2}$. From Lemma~\ref{lem:adversarial LQR sol}, this further implies
$P_{\gamma_2} \prec P_{\gamma_\infty} \prec \gamma_\infty^2 I$. \hfill$\blacksquare$

Therefore, Lemma~\ref{lem: gamma ordering} immediately implies $\norm{P_\gamma} \prec \gamma_\infty^2$, and thus:
\begin{align}\label{eq: upper bound second term}
    \frac{\norm{P_\gamma}^2}{\gamma^2 - \norm{P_\gamma}} \leq \frac{\gamma_\infty^4}{\gamma^2 - \gamma_\infty^2}.
\end{align}
Upper bounding $\norm{P_\gamma - P_\star}$ is a bit more involved. As previewed in Section~\ref{sec: upper bound}, we show that defining 
\begin{align}
    \label{eq:perturbed dynamics}
    \tilde{A} &= \paren{I + \paren{\gamma^2 I - P_\gamma}^{-1} P_\gamma} A, \quad
    \tilde{B} = \paren{I  + \paren{ \gamma^2 I - P_\gamma}^{-1} P_\gamma} B
\end{align}
we can re-write the state update under the adversary as
\begin{align*}
    x_{t+1} &= Ax_t + Bu_t + \delta_t = \tilde{A}x_t + \tilde{B}u_t.
\end{align*}
We then show that the gaps $\norm{\tilde{A} - A},\;\norm{\tilde{B} - B}$ can be upper bounded in terms of $\gamma$ in the following lemma, after which we may adapt some techniques introduced in Proposition~3 from \cite{mania2019certainty} to get an upper bound on $\norm{P_\gamma - P_\star}$.

\begin{lemma}\label{lem: eps to gamma conversion}
Given an arbitrary control sequence $\curly{u_t}_{t\geq 0}$, and an adversarial noiseless LQR instance at level $\gamma > 0$, we recall the adversary $\delta_t$ can be expressed as:
\begin{align*}
    \delta_t &= \paren{\gamma^2 I - P_\gamma }^{-1} P_\gamma \paren{Ax_t + Bu_t}.
\end{align*}
Defining $\tilde{A}, \tilde{B}$ as in Equation~\eqref{eq:perturbed dynamics}, we have the following perturbation bounds 
\begin{align*}
    \norm{\tilde{A} - A} &\leq \frac{\gamma_\infty^2}{\gamma^2 - \gamma_\infty^2 } \norm{A} \\
    \norm{\tilde{B} - B} &\leq \frac{\gamma_\infty^2}{\gamma^2 - \gamma_\infty^2 } \norm{B}.
\end{align*}
Therefore, given a desired level $\zeta > 0$, choosing the adversarial level
\begin{align*}
    \gamma^2 \geq \gamma_\infty^2 + \frac{\gamma_\infty^2}{\zeta}\max\curly{\norm{A}, \norm{B}},
\end{align*}
guarantees $||\tilde{A} - A||, ||\tilde{B} - B|| \leq \zeta$.

\end{lemma}

\textit{Proof:} We can apply rudimentary matrix bounds combined with Lemma~\ref{lem: gamma ordering} to get
\begin{align*}
    \norm{\tilde{A} - A} &= \norm{\paren{P_{\gamma} - \gamma^2 I}^{-1} P_\gamma A} \\
    &\leq \norm{\paren{P_{\gamma} - \gamma^2 I}^{-1}} \norm{P_\gamma} \norm{A} \\
    &\leq \frac{\norm{P_\gamma}}{\gamma^2  - \norm{P_{\gamma}}} \norm{A} \\
    &\leq \frac{\gamma_\infty^2}{\gamma^2 - \gamma_\infty^2} \norm{A},
\end{align*}
which we may repeat for $\norm{\tilde{B} - B}$. Inverting
\[
\frac{\gamma_\infty^2}{\gamma^2 - \gamma_\infty^2} \max\curly{\norm{A},\norm{B}} \leq \zeta,
\]
for $\gamma^2$, we get the desired result. \hfill $\blacksquare$






\begin{proposition}[Upper Bound on $\gamma$-Adversarial versus Nominal DARE Solution]\label{prop: advrobust DARE upper bound ctrlbility}
Let $P_\star$ be the solution to $\dare\paren{A, B, Q, R}$, and $P_\gamma$ be the solution to $\dare\paren{A, \bmat{B & I}, Q,\bmat{R & 0 \\ 0 & -\gamma^2 I}}$, and let $(A,B, Q^{1/2})$ be controllable and detectable. Define $\kappa(Q, R) = \frac{\max\curly{\sigma_{\max}(Q), \sigma_{\max}(R)}}{\min\curly{\sigma_{\min}(Q), \sigma_{\min}(R)}}$,
$\tau(A, \rho) := \sup\curly{\norm{A^k}\rho^{-k}: k \geq 0}$, 
and $\beta = \max\curly{1, \frac{\gamma_\infty^2}{\gamma^2 - \gamma_\infty^2} \tau\paren{A, \rho} + \rho}$, where $\rho > \rho(A)$. Then for
\begin{align*}
    \gamma^2 \geq \gamma_\infty^2 + \frac{3}{2} \ell^{3/2} \beta^{\ell - 1} \sigma_{\min}(W_\ell(A,B))^{-1/2} \tau(A, \rho)^2 \paren{\norm{B} + 1} \max\curly{\norm{A}, \norm{B}}  \gamma_\infty^2,
\end{align*}
the following bound holds:
\begin{align}
    \norm{P_\gamma - P_\star} &\leq 16 \frac{\gamma_\infty^4}{\gamma^2 - \gamma_\infty^2} \ell^{5/2} \beta^{2(\ell - 1)} \paren{1 + \sigma_{\min}(W_\ell(A,B))^{-1/2}} \tau(A, \rho)^3  \paren{\norm{B}+1}^2 \kappa(Q,R)  \norm{P_\star}.
\end{align}

\end{proposition}

The proof of Proposition \ref{prop: advrobust DARE upper bound ctrlbility} is mostly similar to Proposition~3 in \cite{mania2019certainty}. However, we note that our requirement on the size of $\gamma$ is much more lenient than the requirement on $\norm{\tilde{A} -A},\norm{\tilde{B}-B}$ in Proposition 3 of \cite{mania2019certainty}. In short, this is because in our setting we are guaranteed $P_{\gamma} \succ P_\star$, and thus $\norm{P_\gamma - P_\star} = \lambda_{\max}(P_{\gamma} - P_\star)$. Therefore, upper bounding $\lambda_{\max}(P_\gamma - P_\star)$ suffices. Unlike the general case of perturbed DARE solutions, we do not need the gap to be small enough for the upper bound on $\lambda_{\max}(P_\gamma - P)$ to be less than $1/2$. Our only size requirement on $\gamma$ comes from guaranteeing the minimum singular value of the $\ell$-controllability gramian of $(\tilde{A}, \tilde{B})$ to be at least $1/2$ that of the nominal system $(A,B)$. 

Combining Lemma~\ref{lem: controller gap to DARE gap}, Equations~\eqref{eq: upper bound on Sigma(K_gamma)} and~\eqref{eq: upper bound second term}, and Proposition~\ref{prop: advrobust DARE upper bound ctrlbility} together yields the final upper bound on $\NC(K_\gamma) - \NC(K)$.





\subsection{Proof of Theorem~\ref{thm:lower bound}}

We start by stating and proving the spectral upper and lower bounds mentioned in Section~\ref{s: p-r tradeoffs lower bounds}.
\begin{lemma}     \label{lem: spectral UB+LB}
Under the assumption $(A,B,Q^{1/2})$ is stabilizable and detectable, and for $\gamma \geq \tilde{\gamma}_\infty$, we have the following bounds:
    \begin{align*}
        \tilde P_\gamma - P_\gamma 
        &\preceq \norm{K_\gamma-K_\star}^2 \norm{R+B^\top M_\gamma B} W_\infty(\tilde A + \tilde B K_\star, I) \\
        P_\gamma(\gamma^2 I - P_\gamma)^{-1} P_\gamma +  \tilde P_\gamma - P_\star &\succeq \frac{\sigma_{\min}(P_\star)^2}{\gamma^2 - \sigma_{\min}\paren{P_\star}} \paren{W_\infty(A+BK_\star, I)}
    \end{align*}
\end{lemma}

\textit{Proof:} We start with the upper bound. Begin by recalling that for any $x$, $x^\top \tilde P_\gamma x$ is the cost of applying controller $K_\star$ in the noiseless adversarial setting at level $\gamma$ from initial state $x$, while $P_\gamma$ is the cost of applying controller $K_\gamma$ in the noiseless adversarial setting at level $\gamma$ from initial state $x$. If we denote the optimal map from the state to the adversary perturbation in the cases where controller $K_\star$ and $K_\gamma$ are deployed as $\tilde N_\gamma$ and $N_\gamma$ respectively, then we may consider this as the cost gap of applying the controller
\[
    \bmat{u_t \\ \delta_t} = \bmat{K_\star \\ \tilde N_\gamma} x_t
\]
to the system 
\[
    x_{t+1}=Ax_t + \bmat{B & I} \bmat{u_t \\ \delta_t}
\]
with cost $x_t^\top Q x_t + \bmat{u_t \\ \delta_t}^\top \bmat{R & 0\\0 & -\gamma^2 I} \bmat{u_t \\ \delta_t}$. Similarly, we can consider applying the controller
\[
    \bmat{u_t \\ \delta_t} = \bmat{K_\gamma \\ N_\gamma},
\]
which is optimal in this setting. 

Then by Lemma 10 in \cite{fazel2018global}, we have that
\begin{align*}
    x^\top \tilde P_\gamma x - x^\top P_\gamma x = \sum_{t=0} \paren{x_t'}^\top \bmat{K_\star - K_\gamma \\ \tilde N_\gamma - N_\gamma}^\top\paren{\bmat{R & 0\\0 & -\gamma^2 I} + \bmat{B & I}^\top P_\gamma \bmat{B & I}}\bmat{K_\star - K_\gamma \\ \tilde N_\gamma - N_\gamma} x_t'
\end{align*}
where $x_0' = x$, and $x_{t+1}' = (A+BK_\star + \tilde N_\gamma) x_t' $. Next, observe that 
\begin{align*}
    \bmat{R & 0\\0 & -\gamma^2 I} + \bmat{B & I}^\top P_\gamma \bmat{B & I} \preceq \bmat{R+B^\top M_\gamma B & 0 \\ 0 & 0}
\end{align*}
To see that this is the case, we will show that
\begin{align*}
   \bmat{B^\top (P_\gamma - M_\gamma) B & B^\top P_\gamma \\ P_\gamma B & P_\gamma - \gamma^2 I} \preceq 0
\end{align*}
Expanding $M_\gamma = P_\gamma + P_\gamma (\gamma^2 I- P_\gamma)^{-1} P_\gamma$, and multiplying the expression by $-1$, we can see that this is equivalent to showing that
\begin{align*}
   \bmat{B^\top P_\gamma(\gamma^2I - P)^{-1}P_\gamma B & B^\top P_\gamma \\ P_\gamma B & \gamma^2I - P_\gamma} \succeq 0.
\end{align*}
As $P_\gamma \preceq \gamma^2 I$, this follows from a Scur complement check for positive definiteness by the fact that $\gamma^2 I - P_\gamma \succ 0$. Then we have that
\begin{align*}
    x^\top \tilde P_\gamma x - x^\top P_\gamma x \leq  \sum_{t=0} \paren{x_t'}^\top (K_\star - K_\gamma)^\top\paren{R+B^\top M_\gamma B}(K_\star - K_\gamma) x_t'
\end{align*}
or, by the fact thtat $x_{t+1}' = A+BK_\star + \tilde N_\gamma x_t' = \paren{I+(\gamma^2 I-\tilde P_\gamma)^{-1}\tilde P_\gamma}(A+BK_\star) x_t'$,
\begin{align*}
    \tilde P_\gamma - P_\gamma &\preceq \sum_{t=0} \paren{\paren{\paren{I + (\gamma^2 I - \tilde P_\gamma)^{-1} \tilde P_\gamma} (A+BK_\star)}^t}^\top  (K_\star - K_\gamma)^\top \\
    &\quad\quad\quad \paren{R+B^\top M_\gamma B}(K_\star - K_\gamma) \paren{\paren{I + (\gamma^2 I - \tilde P_\gamma)^{-1} \tilde P_\gamma} (A+BK_\star)}^t 
\end{align*}
as we wanted to show.

Moving on to the lower bound, note that we may express 
\begin{align*}
    \tilde P_\gamma &= \dlyap\paren{(I + E) (A+BK_\star), Q + K_\star^\top R K_\star -\gamma^2 (A+BK_\star)^\top E^\top E (A+BK_\star)} \\
    P_\star &= \dlyap \paren{A + BK_\star, Q + K_\star^\top RK_\star} \\
    E &= (\gamma^2 I - \tilde P_\gamma)^{-1} \tilde P_\gamma
\end{align*}
Then 
\begin{align*}
    \tilde P_\gamma - P_\star &= (A+BK_\star)^\top (I+E)^\top \tilde P_\gamma (I+E) (A+BK_\star) - \gamma^2 (A+BK_\star)^\top E^\top E (A+BK_\star)  \\
    & \quad \quad - (A+BK_\star)^\top  P_\star (A+BK_\star) \\
    &= (A+BK_\star)^\top (\tilde P_\gamma-  P_\star) (A + BK_\star)  + (A+BK_\star)^\top \paren{E^\top \tilde P_\gamma + \tilde P_\gamma E + E^\top \tilde P_\gamma E - \gamma^2 E^\top E} (A+BK_\star)
\end{align*}
However, we have that 
\begin{align*}
    E^\top \tilde P_\gamma = \tilde P_\gamma E = \tilde P_\gamma (\gamma^2 I - \tilde P_\gamma) \tilde P_\gamma 
\end{align*}
and 
\begin{align*}
    E^\top \tilde P_\gamma E - \gamma^2 E^\top E =  \paren{\tilde P_\gamma - \gamma^2 I}  E = \tilde P_\gamma^\top (\gamma^2 I - \tilde P_\gamma) \tilde P_\gamma.
\end{align*}
Then 
\[
    \tilde P_\gamma -  P_\star = \dlyap(A+BK_\star, (A+BK_\star)^\top \tilde P_\gamma(\gamma^2 I - \tilde P_\gamma)^{-1} \tilde P_\gamma (A+BK_\star)).
\]

We can expand the Lyapunov solution as an infinite sum:
\[
    \tilde P_\gamma -  P_\star = \sum_{t=1}^\infty \paren{(A+BK_\star)^t}^\top \tilde P_\gamma(\gamma^2 I - \tilde P_\gamma)^{-1} P_\gamma (A+BK_\star)^t
\] 
Next, note that
\[
    \tilde P_\gamma(\gamma^2 I - \tilde P_\gamma)^{-1} \tilde P_\gamma \succeq \frac{\sigma_{\min}(\tilde P_\gamma)^2}{\gamma^2- \sigma_{\min}\paren{\tilde P_\gamma}} I \succeq \frac{\sigma_{\min}(P_\star)}{\gamma^2 - \sigma_{\min}\paren{P_\star}} I
\]  
and thus 
\[
    \tilde P_\gamma -  P \succeq \frac{\sigma_{\min}(P_\star)^2}{\gamma^2 - \sigma_{\min}\paren{P_\star}}\sum_{t=1}^\infty \paren{(A+BK_\star)^t}^\top  (A+BK_\star)^t 
\]
Adding in the term 
\[
    P_\gamma(\gamma^2 I - P_\gamma)^{-1} P_\gamma \succeq \frac{\sigma_{\min}( P_\gamma)^2}{\gamma^2- \sigma_{\min}\paren{ P_\gamma}} I \succeq \frac{\sigma_{\min}(P_\star)}{\gamma^2 - \sigma_{\min}\paren{P_\star}} I
\]
allows us to start the sum from $0$, providing the desired result. \hfill $\blacksquare$

\begin{lemma}
    \label{lem:controller gap LB}
    As long as $\gamma$ satisfies
    \begin{align*}
        \gamma^2 &\geq \sigma_{\min}(P_\star) + \sigma_{\min}(P_\star)^2 \sigma_{\min}(A+BK_\star)^2 \frac{1}{2 \norm{R+B^\top M_\gamma B}} \\
        &\quad \quad \quad \norm{B^\top W_\infty\paren{\paren{I + (\gamma^2 I - \tilde P_\gamma)^{-1} \tilde P_\gamma}(A+BK_\star), I}}
        \norm{B^\top W_\infty(A+BK_\star, I) },
    \end{align*}
    we have the following bound
    \begin{align*}
        \norm{K_\gamma - K_\star} \geq \frac{\sigma_{\min}(P_\star)^2 \norm{B^\top W_\infty(A+BK_\star,I)} \sigma_{\min}(A+BK_\star) }{2\norm{R + B^\top  M_\gamma B} (\gamma^2 - \sigma_{\min}\paren{P_\star})}
    \end{align*}
\end{lemma}

\textit{Proof:}
Beginning with the lower bound in \eqref{eq:lb on controller gap} and applying Lemma~\ref{lem: spectral UB+LB}, we find
\begin{align*}
    \norm{K_\gamma-K_\star} &\geq  \frac{\sigma_{\min}(A+BK_\star)}{\norm{R + B M_\gamma B}} \bigg(\frac{\sigma_{\min}(P_\star)}{\gamma^2 - \sigma_{\min}(P_\star)}\norm{B^\top W_\infty(A+BK_\star, I)  } \\
    &\quad\quad\quad -\norm{B^\top W_\infty\paren{\paren{I + (\gamma^2 I - \tilde P_\gamma)^{-1} \tilde P_\gamma} (A+BK_\star), I} } \norm{K_\gamma-K_\star}^2 \norm{R+B^\top M_\gamma B}\bigg)
\end{align*}

In particular, by defining 
\begin{align*}
    a &= \sigma_{\min}(A+BK_\star) \norm{B^\top W_\infty\paren{\paren{I + (\gamma^2 I - \tilde P_\gamma)^{-1} \tilde P_\gamma} (A+BK_\star), I}}  \\
    c &= \frac{\sigma_{\min}(P_\star)^2}{\gamma^2 - \sigma_{\min}(P_\star)}\frac{\norm{B^\top W_\infty(A+BK_\star,I)}}{\norm{R + B^\top M_\gamma B}} \sigma_{\min}(A+BK_\star),
\end{align*}
the above bound can be written
\begin{align*}
    a\norm{K_\gamma-K_\star}^2 + \norm{K_\gamma-K_\star} -c \geq 0
\end{align*}
which, considering that $\norm{K_\gamma-K_\star} \geq 0$, implies that
\begin{align*}
    a\norm{K_\gamma-K_\star} \geq \sqrt{\frac{1}{4}  + ac} - \frac{1}{2}.
\end{align*}
Under the assumption that $ac \leq 2$, this may be simplified to $\norm{K_\gamma-K_\star} \geq \frac{1}{2} c$. $\blacksquare$

Theorem~\ref{thm:lower bound} then follows from the general lower bound in equation \eqref{eq:cost gap lower bound} and the bound from Lemma~\ref{lem:controller gap LB}.

\end{document}